\documentclass[usenatbib,useAMS]{mn2e}
\usepackage{graphicx,portland,lscape}

\title[On the natural sub-division of galaxies]{The Millennium Galaxy Catalogue: On the natural sub-division of galaxies} 

\author[Ellis et al.\ ]{S.C. Ellis$^{1}$\thanks{E-mail: sce@aao.gov.au}, S.P. Driver$^{2}$, P.D. Allen$^{2}$, J. Liske$^{3}$, J. Bland-Hawthorn$^{1}$,  \newauthor R. De Propris$^{4}$ 
 \\ 
$^{1}$ Anglo-Australian Observatory, P.O. Box 296, Epping, NSW 2121, Australia\\
 $^{2}$ Research School of Astronomy and Astrophysics, Mount Stromlo Observatory,
Cotter Road, Weston, ACT 2611, Australia \\
$^{3}$ European Southern Observatory, Karl-Schwarzschild-Stra{\ss}e 2, 85748 Garching b.\ M{\"{u}}nchen, Germany\\
$^{4}$ H.\ H.\ Wills Physics Lab., University of Bristol, Tyndall Ave., Bristol, BS8 1TL, UK}

\date{Accepted...... Received .....}

\begin{document}

\label{firstpage}

\maketitle

\begin{abstract}
The distribution of global photometric, spectroscopic, structural and
morphological parameters for a well defined sample of 350 nearby
galaxies has been examined.  The usual trends were recovered
demonstrating that E/S0 galaxies are redder, more quiescent, more
centrally concentrated and possess larger S{\'{e}}rsic indices than
later type galaxies.

Multivariate statistical analyses were performed to examine the
distribution of all parameters simultaneously.  The main result of
these analyses was the existence of only two classes of galaxies,
corresponding closely to early and late types.  Linear discriminant
analysis was able to reproduce the classifications of early and late
types galaxies with high success, but further refinement of galaxy
types was not reproduced in the distribution of observed galaxy
properties.  A principal components analysis showed that the major
variance of the parameter set corresponded to a distinction between
early and late types, highlighting the importance of the distinction.
A hierarchical clustering analysis revealed only two clear natural
classes within the parameter set, closely corresponding to early and
late types.  Early and late types are clearly distinct and the
distinction is of fundamental importance.  In contrast, late types
from Sa to Irr are smoothly distributed throughout the parameter
space.  

A population of galaxies classified by eye as elliptical/lenticular,
and exhibiting concentration indices similar to early-types were found
to have a significant star-formation activity.  These galaxies are
preferentially faint, suggesting they are low-mass systems.

\end{abstract}


\section{Introduction}

Galaxies display a large variety of observational and physical
properties.  Some properties, such as colour, emission-line strength
and far-infrared luminosity are well correlated with morphological
appearance such as bulge-to-disc ratio, tightness of the spiral arms
etc.\ (\citealt{rob94}; \citealt{bin98}).  These correlations are
primarily due to the amount of star-formation, gas and dust present in
the galaxies (\citealt{ken98}).  Other properties such as luminosity
and size, correlate well with galaxy mass, but have weaker
correlations with morphology, since galaxies of different morphology
have a large range in mass.

Understanding these relationships in detail is paramount to
constructing a consistent picture of galaxy evolution.  Galaxy
properties are observed to evolve both as a function of environment
and age (e.g.\ \citealt{bel04}).  In order to have a clear
understanding of how the observed evolution of each property relates
to the evolution of other properties, it is first necessary to
understand how the properties are related to one another.

For example, many galaxy properties are observed to be dependent on
the density of the galaxy's environment (\citealt{bal04};
\citealt{hog04}).  The morphological mix of galaxies changes as a
function of local projected galaxy density, and cluster-centric
radius, with the fraction of ellipticals and lenticulars increasing at
higher densities, and a corresponding decrease in the fraction of
spirals (\citealt{dre80}; \citealt{whi93}).  This in turn leads to
differences in the luminosity functions of galaxies in different
density regimes (e.g.\ \citealt{cro05}).  The star-formation rate of
galaxies declines with increasing density at densities greater than a
specific threshold (\citealt{lew02}; \citealt{bal04}).  The average
colour of galaxies becomes redder in high density environments
(\citealt{dep05b}), again with an apparent density threshold
(\citealt{kod01}).  Understanding whether these density relations are
manifestations of the same physical change in the galaxies, or whether
several mechanisms are responsible, requires understanding the
relations between the morphology, colour and star-formation.

Some galaxy properties evolve with redshift, in a manner dependent on
density.  At $z\approx 0.5$ the morphology-density relation shows an
increase in the fraction of spirals in high density environments
(\citealt{dre97}).  Similarly the fraction of blue galaxies in
clusters increases with redshift (\citealt{but84}).
Disentangling the connected effects of galaxy age and environment is
difficult, since transformations of galaxies due to infall into dense
environments could produce similar observational results to a
difference in galaxy age as a function of environment.  It is crucial
to understand the relationships between the observational properties
of galaxies at low redshift in order to make progress in understanding
galaxy evolution.

Many of these correlations are already well known and understood.  For
instance \citet{hub36} introduced the relation between galaxy colour
and morphology.  Some correlations are more recent, such as the
correlation between star formation activity and the $\eta$ parameter
of the 2dF Galaxy Redshift Survey (2dFGRS), introduced by
\citet{mad02}, and based on a principal components analysis of 2dFGRS
spectra.

We present here a detailed analysis of the distribution and
correlations of the observational properties of a well-defined local
sample of galaxies for which we have compiled global photometric,
spectroscopic and structural parameters.  We expand on previous work
regarding correlations between galaxy properties in two major ways.
Firstly, we simultaneously examine correlations between a large number
of parameters using multivariate statistical analysis techniques.
Secondly, we do not limit ourselves to correlations with \emph{a
priori} assigned morphological type (discrimination), though these are
still investigated and form a major part of the work, but we also
search the parameter space for natural clustering of galaxies
(classification).

The paper is organised as follows.  After introducing the sample 
and describing the catalogue in
Section~\ref{sec:sample}, we describe the characteristics of a
population of blue spheroidal galaxies in Section~\ref{sec:de}.
Following this we examine the distributions of each parameter as a
function of morphological type in Section~\ref{sec:correl}, in order
to gauge the overall trends displayed by the galaxies.  In
Section~\ref{sec:pca} the parameter set is more thoroughly explored
using a principal components analysis to determine combinations of
parameters which account for a large amount of the total variance of
the parameter set.  The distribution of the first principal component
separates early and late galaxies with good success.
Section~\ref{sec:lda} further examines the separation of galaxies of
different morphological type using linear combinations of parameters
via a linear discriminant analysis.  It is found that early type
galaxies can be identified with good success, but discrimination of
other types is less successful.  In Section~\ref{sec:hca} a
hierarchical clustering analysis is performed to look for distinct
groups of galaxies within the parameter set, without any prior input
regarding morphological classifications.  Only two `natural' classes
of galaxies are found, corresponding very well with early and late
type galaxies.  The implications of the work are discussed in
Section~\ref{sec:discuss}.

A future paper will extend this work by exploring the results of a
bulge-disc decomposition as additional input to the analysis.

\section{Sample}
\label{sec:sample}

The Millennium Galaxy
Catalogue\footnote{http://www.eso.org/$\sim$jliske/mgc} (MGC) is a
wide, medium-deep, $B$-band imaging survey obtained with the Wide
Field Camera on the Isaac Newton Telescope on La Palma
(\citealt{lis03}).  It covers 37.5 square degrees with magnitudes in
the range $13 < B < 24$ mag.  It is $\approx 96\%$ spectroscopically
complete to $B<20$ mag (see \citealt{dri05}).

In total MGC-BRIGHT (i.e.\ $B<20$ mag) contains 10095 galaxies.
Although a comparatively small catalogue in comparison to the Two
Degree Field Galaxy Redshift Survey (2dFGRS; $\sim 250,000$ galaxies,
\citealt{col01}) and the Sloan Digital Sky Survey (SDSS; data release
three $\sim 370,000$ galaxies, \citealt{aba05}), the MGC has several
advantages that make it ideal for studying galaxy classification.  The
depth of the MGC redshift survey surpasses both the 2dFGRS ($b_{{\rm j}}=19.45$ mag)
and the SDSS ($r=17.8\approx {\rm B}=18.8$ mag) spectroscopic limits.  This increase in
sensitivity allows the study of intrinsically lower luminosity
galaxies than the larger surveys.  This in turn allows the construction of a
more representative sample of galaxies, and the investigation of the
dependence of galaxy properties with absolute magnitude.

A further important advantage is the faint surface brightness limit of
the MGC ($\mu_{{\rm B}}=26$ mag arcsec$^{-2}$).  \citet{dri05} examine
the distribution of galaxies as a function of magnitude and surface
brightness.  It is shown that the inclusion or omission of low surface
brightness galaxies can have a significant effect on the resulting
luminosity function.  The presence of low surface brightness galaxies
again allows construction of more representative catalogues.  The
biases resulting from a deficit of low surface brightness galaxies may
be significantly different from biases arising from magnitude
selection, since galaxies may be relatively luminous whilst having low
surface brightness.

Finally the quality of the images in the MGC are a major asset for
studying galaxy classification.  The median seeing of the survey of
1.3 arcseconds (\citealt{lis03}), combined with the surface brightness
limit of $\mu_{{\rm B}}=26$ mag arcsec$^{-2}$, produce very high
quality images for a survey of this scale.  The good image
quality allows the structural properties of the galaxies to be
quantified, through surface brightness fitting with {\sc gim2d} (Allen
et al., in preparation) and visual inspection (Driver et al., in
preparation), allowing this important aspect of galaxy classification
to be investigated.

\subsection{The catalogue}
\label{sec:pars}

We have acquired a large set of global parameters to thoroughly investigate the
characterstics of galaxies within the catalogue.  The list of all
parameters used in the analysis is given in Table~\ref{tab:parlist}.
A description, and where necessary the derivation, of each parameter
is given below.

\begin{table*}
\caption{List of all parameters used in the analysis}
\label{tab:parlist}
\begin{tabular}{l}
Parameter \\ \hline
Redshift\\            
MGC $B$ band absolute Kron mag., dust, K \& E-corrected (Vega mag)   \\
SDSS $u-r$ absolute Petrosian colour, dust, K \& E-corrected (AB mag)  \\ 
MGC $B$ band absolute effective surface brightness, dust, K \& E-corrected (Vega mag) \\  
2dFGRS $\eta$ parameter\\           
GIM2D S{\'{e}}rsic index\\ 
Concentration index (inner 10\%)\\         
Asymmetry index (background corrected)     (inner 10\%)\\       
SCE's visual classification\\
SPD's visual classification          
\end{tabular}
\end{table*}

\subsubsection{Morphological classification}
\label{sec:morph}

The 350 galaxies selected for study (see Section~\ref{sec:subsamp}) have been visually inspected and
classified into broad morphological classes: E/S0, Sabc and Sd/Irr.
Two different techniques were employed by different authors to
classify the galaxies, which makes it possible to assess the accuracy
of the classifications through comparison of the results.  SPD used
images of the galaxies printed at several surface brightness levels.
SCE used purpose written software to display the galaxies alongside
contours of surface brightness.  


Comparison of the classifications of galaxies by different authors is
made in Table~\ref{tab:class} for the 326 galaxies classified by SPD
(with $B \le 19$).  There is $\approx 80\%$ agreement on E/S0 and Sabc
galaxies, but somewhat less on Sd/Irrs.  The figure of $\sim 80\%$ may
be regarded as a benchmark, against which to test other classification
schemes.  Henceforth we use the classification of SCE.

\begin{table}
\caption{Number of galaxies classified into each type by SPD and the number of those which were agreed on by SCE.  The percentage of agreement is
also listed}
\label{tab:class}
\begin{tabular}{lccc}
& E/S0 & Sabc & Sd/Irr\\ \hline
SPD & 75 & 179 & 72 \\
SCE & 62 & 140 & 44 \\
\% agreed & 83 & 78 & 61
\end{tabular}
\end{table}

\subsubsection{2dFGRS $\eta$-parameter}
The $\eta$ parameters are obtained from the final 2dFGRS (\citealt{col01}) and are described in
detail by \citet{mad02}. In brief $\eta$ is derived from a
principal components analysis of the 2dF galaxy spectra and is a linear
combination of PC1 and PC2. PC1 contains information from emission and absorption line strength and the continuum in roughly equal amounts, whilst PC2 is dominated by emission and absorption line strengths.  The
linear combination is tuned to maximise the line features and is hence a likely indicator of the current and
previous star-formation rate. The matching of the MGC to the final
2dFGRS catalogue is described in \citet{cros04}. No errors are
provided for $\eta$.

\subsubsection{Rest SDSS-DR1 Petrosian $(u-r)$ colours}
\label{kcorr}
Petrosian $ugriz$ magnitudes are obtained from the SDSS-DR1 catalogue
(see \citealt{aba03}), which overlaps fully with the
MGC. Details of the matching algorithm are given in \citet{dri05}, as is the conversion from observed to rest-frame colour. In
brief the $uBgriz$ colours are used to determine the optimal spectral
template from the spectral library of \citet{pog98}. The optimal
spectra are then used to derive the appropriate K-correction for each
galaxy through each filter. Rest-frame colours are derived from the
rest-frame absolute magnitudes (see Section~\ref{absmag}). The random
error in the final colours are $\Delta$ rest-$(u-r) \approx 0.09$ mag
and are dominated by the uncertainty in the evolutionary correction.

\subsubsection{Absolute magnitude and surface brightness}
\label{absmag}
The absolute $uBgriz$ magnitudes and the absolute $B$-band effective
surface brightness, are derived as stated in \citet{dri05}. This
includes individual K-corrections (as in Section~\ref{kcorr}), a
universal evolutionary correction ($e(z)=2.5 \log_{10}
[(1+z)^{-\beta}]$, $\beta=1.5,0.75,0.75,0.5,0.25,0.13$ for $uBgriz$
respectively), and the standard luminosity distance corrections
(assuming $\Omega_M=0.3, \Omega_{\Lambda}=0.7$ and
$H_{0}=100$km s$^{-1}$ Mpc$^{-1}$). The $B$-band effective surface brightness is derived as: $\mu_{B} =
  m+2.5\log_{10}(2\pi R_{{\rm e}}^2)-10\log_{10}(1+z)-k(z)-e(z)$, where: $m$ is
  the {\sc SExtractor} measured \citet{kro80} magnitude, and $R_{{\rm e}}$ is the half-light
  semi-major axis radius. The latter is derived empirically by growing
  the {\sc SExtractor} specified ellipse until it encloses half the Kron
  flux ($R_{{\rm e}}^{\rm obs}$) and correcting this value for the impact of
  the seeing FWHM ($\Gamma$) according to: $R_{{\rm e}}=\sqrt{(R_{{\rm e}}^{obs})^2-0.32
  \Gamma^2}$ (see \citealt{dri05} for details). The terms $e(z)$ and $k(z)$
  are the evolution and K-corrections as defined above. The random
  error in the absolute $B$-band magnitudes is $0.06$ mag and in the
  absolute effective surface brightness is $0.2$ mag arcsec$^{-2}$.

\subsubsection{S\'ersic profiling}
The S\'ersic profiles are measured using {\sc gim2d} (\citealt{sim98}).
{\sc gim2d} is a 2D image analysis code which requires the raw data image, a
segmentation image (describing which pixels are associated with the
object in question), and the point-spread function (modelled for each
galaxy individually). The application of {\sc gim2d} to the MGC is described
in detail in Allen et al (in preparation). The reproducible accuracy of
the S\'ersic fits have been assessed via repeat observations of $\sim
700$ galaxies and the random error in $\log_{10} n$, where $n$ is the
S\'ersic index, is $<10\%$. The concentration and asymmetry parameters
are also derived via {\sc gim2d} using the definition of \citet{abr96}
with $\alpha=0.1$.

\subsection{Volume selected sub-sample}
\label{sec:subsamp}

For the purposes of investigating the discrimination and
classification of galaxies into different types a sub-sample of the
MGC was culled.  This sample was selected with the intention of
minimising bias in the properties of the galaxies selected whilst
having a large range in magnitude, and exploring a large parameter set
of galaxy properties including spectral, photometric and structural
information.


There is some conflict in trying to explore a large parameter set of
galaxy properties whilst keeping selection bias at a minimum.  Certain
useful galaxy properties such as the 2dFGRS $\eta$ parameter
(hereafter referred to simply as $\eta$) are magnitude limited, since
the survey from which it was derived has a limit of $b_{{\rm
j}}=19.45$ mag.  Also derivation of parameters such as concentration
and asymmetry will likely be less robust for faint galaxies.  Ideally
a volume limited sub-sample would be culled containing galaxies which
have robust measurements of all the relevant properties.  However, due
to the heterogeneous origins of some of the parameters, this is not
possible without compromising the overall size, and hence the
representation, of the sample.  We emphasise, however, that although
different parameters originate from different surveys, any individual
parameter has homogeneous origins, making the distribution of each
parameter very reliable.

\begin{figure}
\centering \includegraphics[scale=0.3,angle=0]{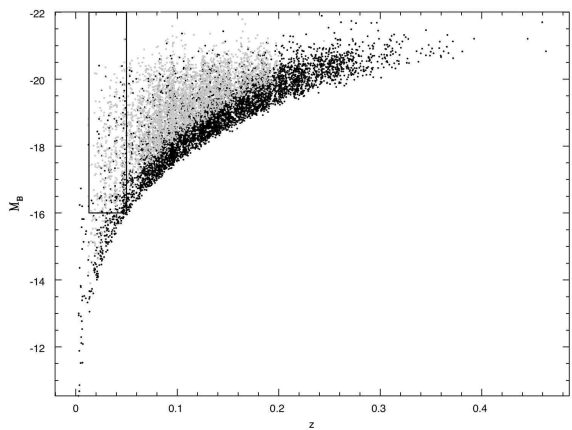}
\caption{Absolute B band magnitude versus redshift for the MGC.  Grey points denote galaxies for which the
2dFGRS $\eta$ parameter is known.  The sub-sample culled consists of the grey points within the box shown,
yielding 350 galaxies.}
\label{fig:sample}
\end{figure}

The sub-sample we selected comprised all galaxies for which $\eta$ has
been measured within $-16 \ge M_{{\rm B}} \ge -22$ mag and $0.013 \le
z \le 0.05$, yielding a total of 350 galaxies.  The sample is
illustrated in Figure~\ref{fig:sample}, and the catalogue is available
from the MGC web
pages.
Note that the sample is not truly volume limited since there is an
apparent magnitude limit of $b_{{\rm j}}<19.45$ mag, set by the
exclusion of galaxies without $\eta$, and furthermore there are surface
brightness and size selection effects intrinsic to the MGC of
$\mu_{{\rm B_{{\rm eff}}}}<25.25$ mag arcsec$^{-2}$ and
$1.06<$half-light radius$<15$ arcsecs (see \citealt{dri05} for full details).


We have looked for the presence of bias in the sub-sample, which may
have been introduced by selecting only those galaxies for which $\eta$
is known.  We have examined the distributions of $M_{{\rm B}}$, $\mu_{{\rm B}}$, $u-r$
and S{\'{e}}rsic index for galaxies with and without $\eta$ within the
same volume.  These parameters are relatively easy to interpret and
should be fairly robust.  The means and standard deviations of the
distributions are given in Table~\ref{tab:bias}.  Also listed are the
biweight location and scatter estimates, which may give a truer
picture of the average and scatter for a skewed distribution (see
\citealt{bee90}).

A Kolmogorov-Smirnov test was performed on each pair of distributions
(i.e.\ those with and without $\eta$) of each parameter, to test if
they are consistent with being drawn from the same population.  The
significance of each test is given in Table~\ref{tab:bias}, where a
low probability indicates that it is unlikely that the two sample are
drawn from the same population.

\begin{figure}
\centering \includegraphics[scale=0.3,angle=270]{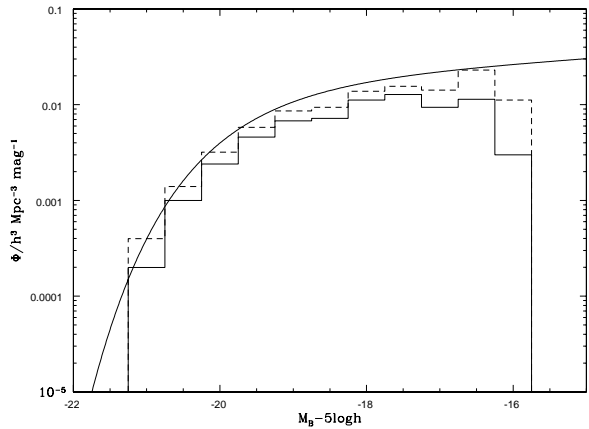}
\caption{The luminosity function of galaxies in the selected sub-sample.  The solid histogram shows only those galaxies for which $\eta$ is known, while the dashed histogram shows the luminosity function for all galaxies within the selected volume.  Overlaid is the best fitting \protect\citet{sch76} function for the whole MGC (\protect\citealt{dri05}).}
\label{fig:lf}
\end{figure}
\begin{figure}
\begin{minipage}[c]{0.25\textwidth}
\centering \includegraphics[scale=0.3,angle=270]{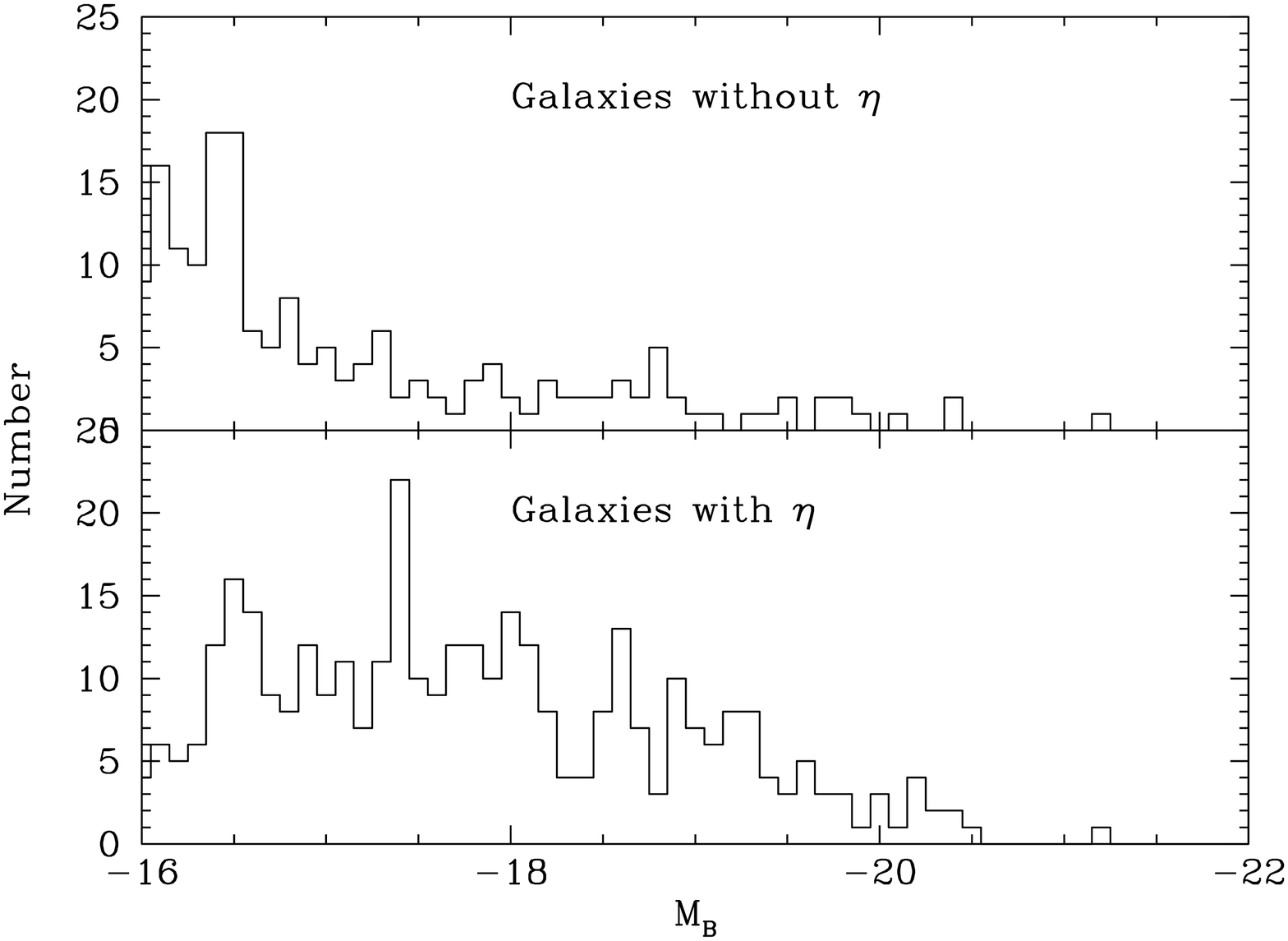}
\end{minipage}
\begin{minipage}[c]{0.25\textwidth}
\centering \includegraphics[scale=0.3,angle=270]{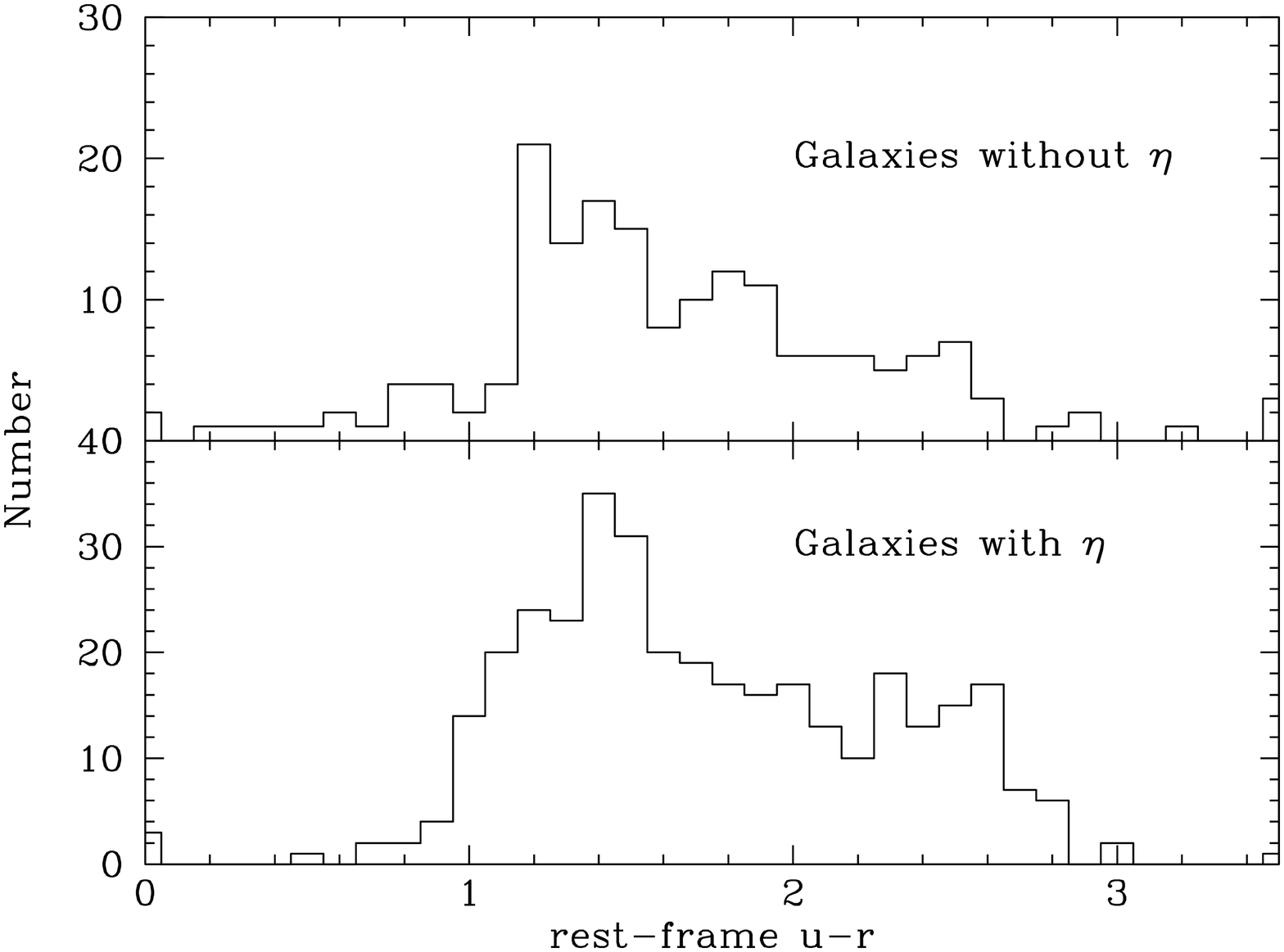}
\end{minipage}
\begin{minipage}[c]{0.25\textwidth}
\centering \includegraphics[scale=0.3,angle=270]{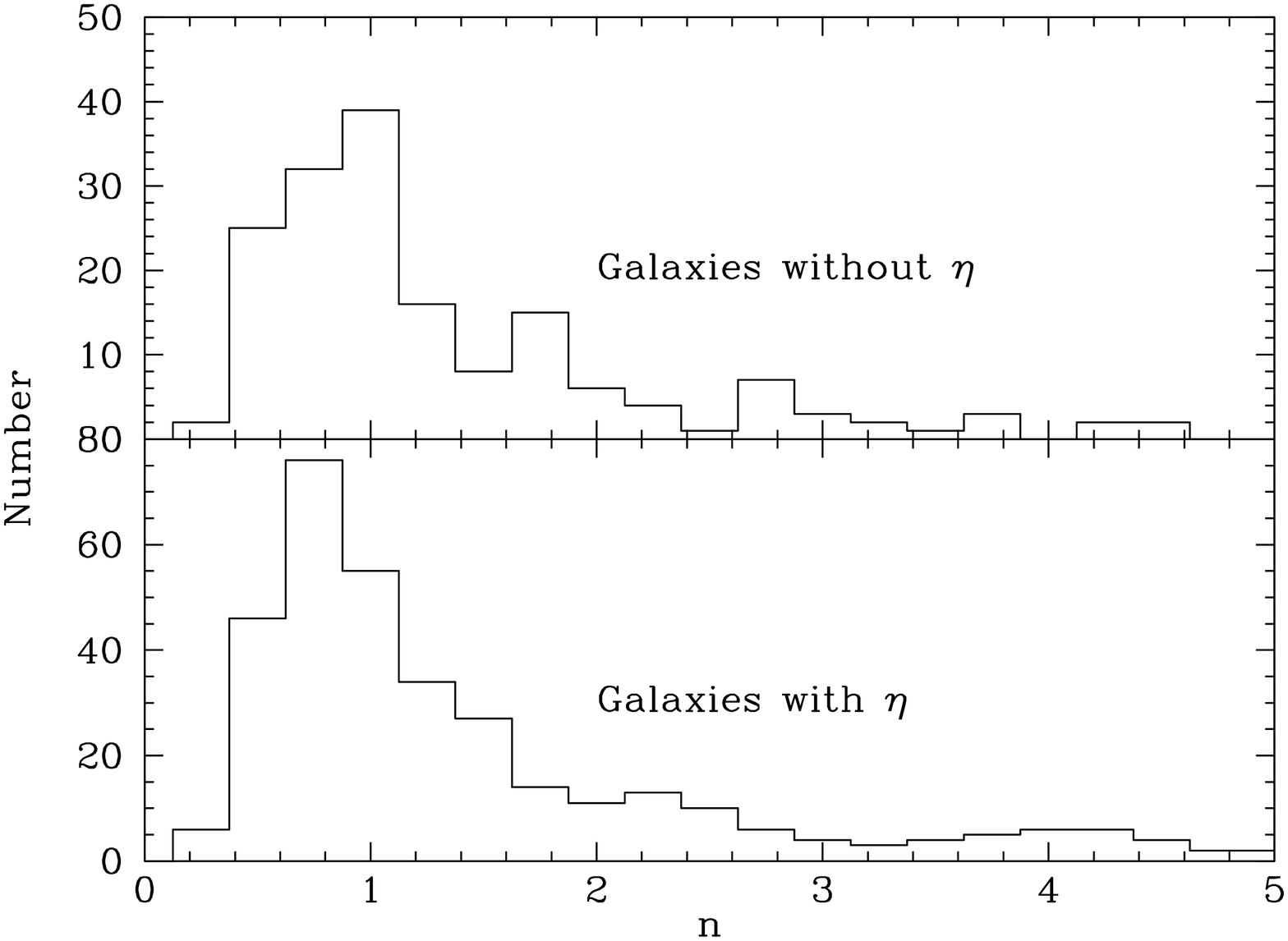}
\end{minipage}
\caption{Distributions of $M_{{\rm B}}$, rest-frame $u-r$ colour and S{\'{e}}rsic index for galaxies with and without $\eta$.  The galaxies without $\eta$ are preferentially faint, but have similar distributions of colour and S{\'{e}}rsic index.}
\label{fig:bias}
\end{figure}

The luminosity function of galaxies in the sub-sample is shown in
Figure~\ref{fig:lf}.  The solid histogram shows only those galaxies
for which $\eta$ is known, while the dashed histogram shows the
luminosity function for all galaxies within the selected volume.
Overlaid is the best fitting \cite{sch76} function for the whole MGC
(\citealt{dri05}).  The luminosity function for the volume selected
sample is consistent with that of \citet{dri05}, indicating that the
volume is representative of the MGC as a whole. The luminosity
function for galaxies with $\eta$ has a lower normalisation due to the
fact that some galaxies have been excluded, but the shape is
consistent, except for a deficit of galaxies in the faintest bin.
This indicates that the selected sub-sample is broadly representative
of the whole MGC, except at the faintest magnitudes where there are
too few galaxies, because the 2dFGRS, from which $\eta$ has been
taken, only reaches magnitudes of $b_{{\rm j}}$=19.45 mag.  This can
also be seen in the top panel of Figure~\ref{fig:bias}, which compares
the distribution of galaxies for which $\eta$ is known to that of the
excluded galaxies; the excluded galaxies are generally fainter.  A
Kolmogorov-Smirnov test shows that the probability is extremely small
that the samples with and without $\eta$ are equivalent.

The middle panel of Figure~\ref{fig:bias} shows the distribution of
rest-frame $u-r$ colour for galaxies with and without $\eta$.  The
distributions appear very similar.  This is reflected in the
statistics listed in Table~\ref{tab:bias}, and a Kolmogorov-Smirnov
test shows that it is reasonable to assume that the two samples are
drawn from the same parent distribution.

The bottom panel of Figure~\ref{fig:bias} shows the distribution of
S{\'{e}}rsic index for the two groups (note there are 2 galaxies fewer
in the group without $\eta$ since the surface brightness fitting
produced spurious results for these galaxies).  The distributions of
both groups are very similar.  Again the statistics are given in
Table~\ref{tab:bias}, and a Kolmgorov-Smirnov test shows that it is
reasonable to assume that the two samples are drawn from the same
parent distribution.

\begin{figure*}
\begin{minipage}[c]{0.45\textwidth}
\centering \includegraphics[scale=0.75,angle=0]{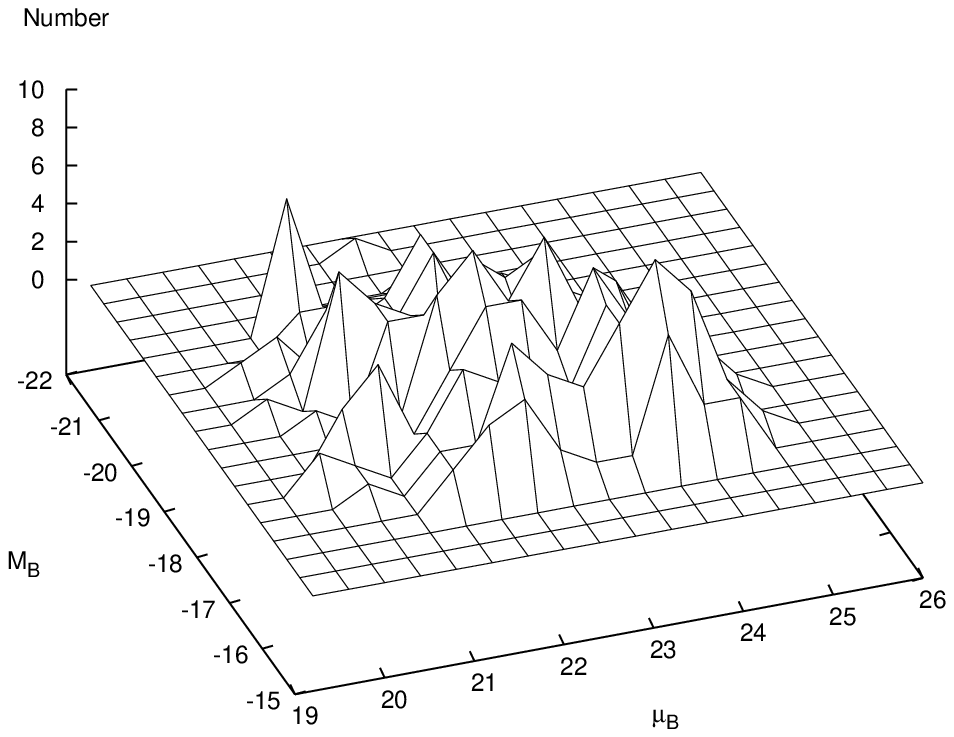}
\end{minipage} %
\begin{minipage}[c]{0.45\textwidth}
\centering \includegraphics[scale=0.75,angle=0]{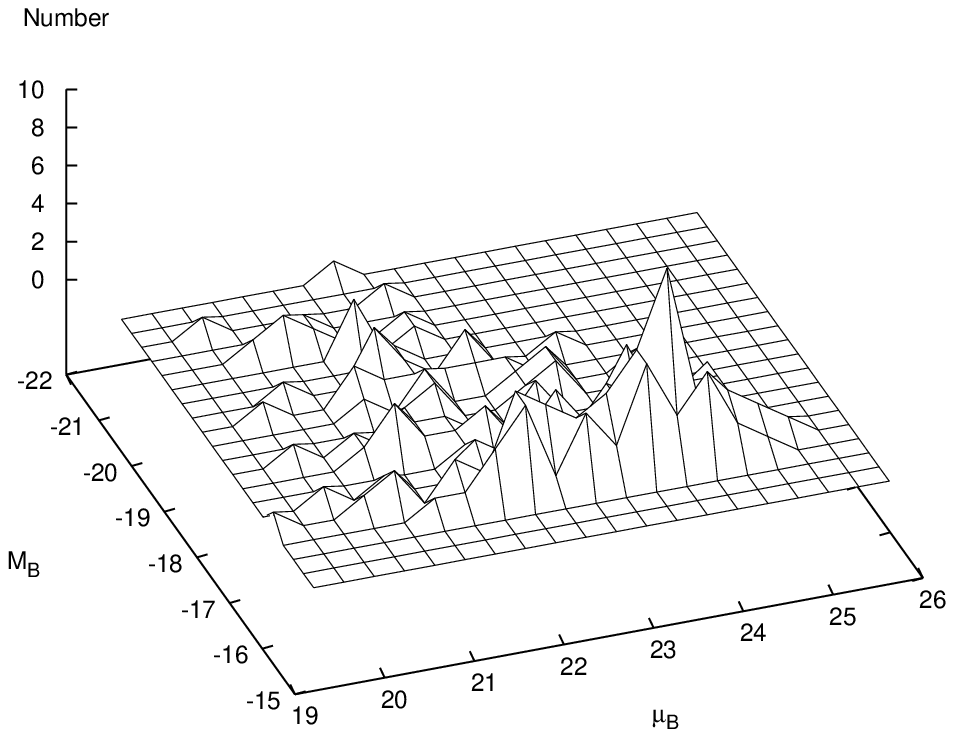}
\end{minipage}
\caption{Histograms of $M_{{\rm B}}$ vs.\ $\mu_{{\rm B}}$ for galaxies with $\eta$ (left) and without (right).}
\label{fig:bbd}
\end{figure*}

We also checked and found no evidence for bias in the surface
brightness of the galaxies with and without $\eta$.  This is at first surprising given that we are biassed against faint galaxies.  However, whilst many of the excluded galaxies are faint and  low surface brightness, there is a population of bright but low surface brightness galaxies which are included and hence the surface brightness distributions appear similar.  Figure~\ref{fig:bbd} shows histograms of the bivariate brightness distributions of the galaxies with and without $\eta$.

It is expected that the colours and S{\'{e}}rsic indices of galaxies
will approximately indicate the type of galaxy and underlying stellar
populations.  The lack of any significant bias in these parameters due
to the exclusion of galaxies without $\eta$ suggests that the
sub-sample culled provides a representative sample of galaxies within
the magnitude range covered.  Caution must be exercised in the
interpretation of any trends in galaxy properties with magnitude
however, since it is clear that the sample contains a deficit of faint
galaxies.

\begin{table*}
\caption{Means, standard deviations, biweight average ($c_{{\rm bi}}$) and scatter($s_{{\rm bi}}$), and Kolmogorov-Smirnov test significance for
galaxies with and without $\eta$.}
\label{tab:bias}
\begin{tabular}{lccccccccr}
& \multicolumn{4}{c}{With $\eta$ (350 galaxies)}&\multicolumn{4}{c}{Without $\eta$ (177 galaxies)} & K-S significance\\
& Mean & $\sigma$ & $c_{{\rm bi}}$ & $s_{{\rm bi}}$ & Mean & $\sigma$ & $c_{{\rm bi}}$ & $s_{{\rm bi}}$ &\\ \hline
$M_{{\rm B}}$ &-17.81 & 1.10 &-17.68 &1.12&-17.15 &1.14&-16.67 & 0.98 & $2.8\times 10^{-13}$\\
rest frame $u-r$ &1.74 & 0.57 &1.70 & 0.57& 1.67 & 1.07 &1.59 &0.56 & 0.24\\
S{\'{e}}rsic index & 1.70 & 2.02 & 1.07 & 0.74 & 1.58 & 1.54 & 1.06 & 0.68 & 0.88 \\
$\mu_{{\rm B}}$ & 22.4 & 1.1 & 22.4 & 1.1 & 22.5 & 1.2 & 22.6 & 1.3 & 0.20 
\end{tabular}
\end{table*}

\section{Blue spheroids} 
\label{sec:de}

\begin{figure}
\centering \includegraphics[scale=0.3,angle=270]{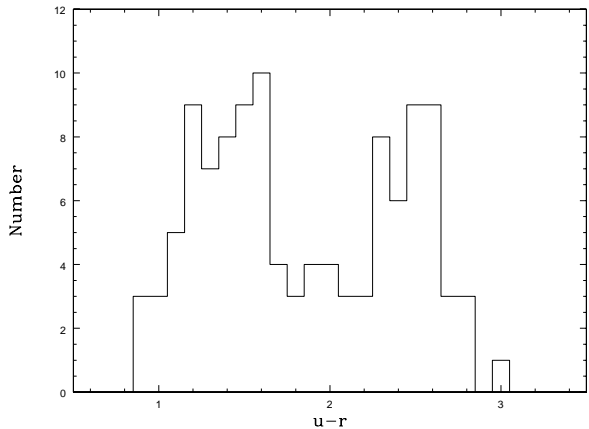}
\caption{The distribution of $u-r$ colours for galaxies classified by eye as E/S0.  The distribution is clearly bimodal.}
\label{fig:dE_col}
\end{figure}

A histogram of the rest-frame $u-r$ colours of the
elliptical/lenticular galaxies is shown in Figure~\ref{fig:dE_col}.
The distribution is clearly bimodal, with $\approx 60\%$
displaying unusually blue colours for conventional ellipticals.  This
is strongly suggestive that there are two distinct populations of
galaxies within the E/S0 class.  This may be due to misclassification
of the galaxies when assigning morphologies by eye, or it may be
revealing a fourth category of galaxies external to the broad classes
of E/S0, Sabc and Sd/Irr. However it is worth noting that if due to
misclassification it occurred independently in both SCEs and SPDs
samples. In order to ascertain the nature of the blue `ellipticals' we
have examined the distribution of several key observational quantities
of the galaxies.

\begin{figure*}
\centering \includegraphics[scale=0.8,angle=0]{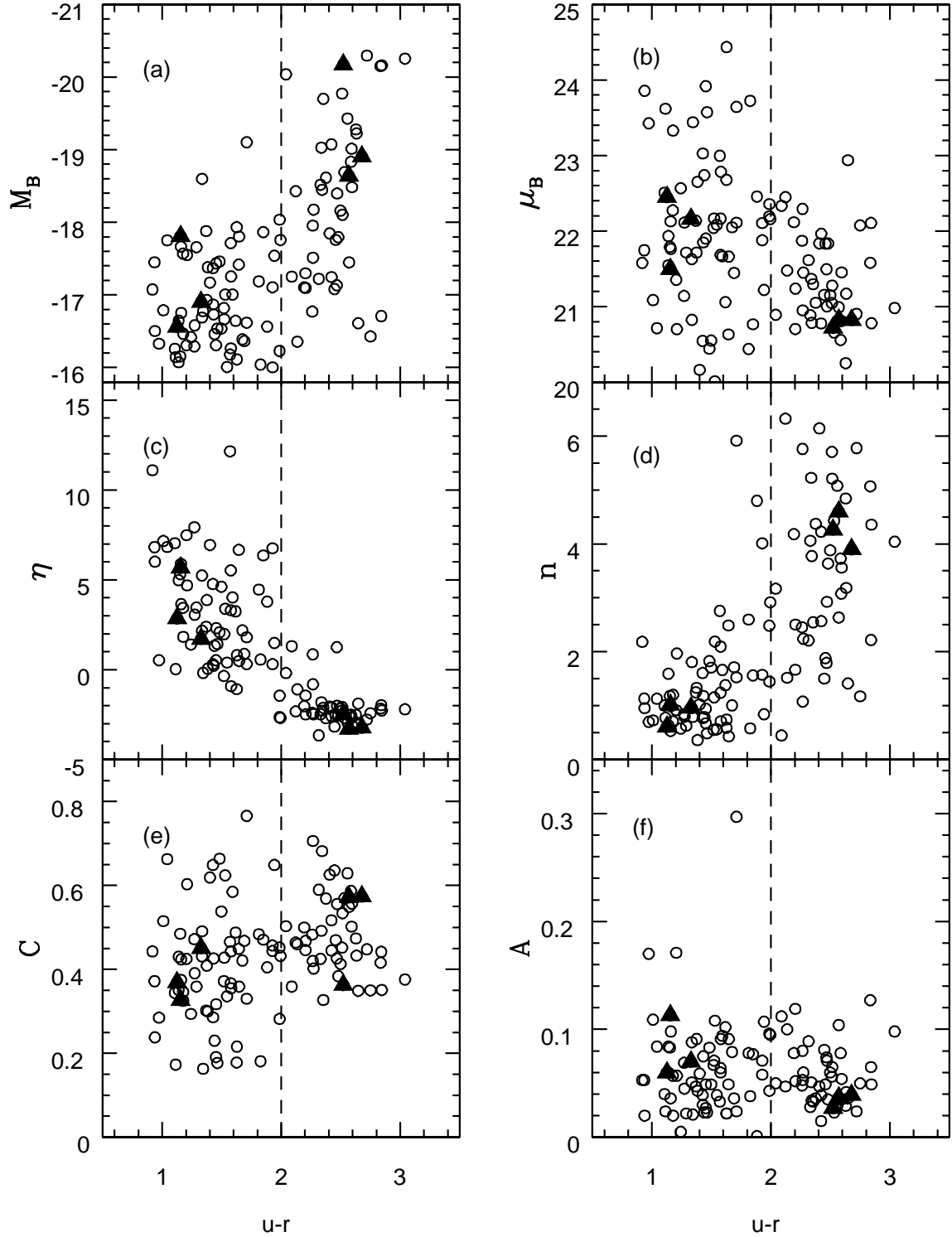}
\caption{The properties of galaxies classified as E/S0 as a function of colour.  Blue E/S0 galaxies are preferentially faint, have higher $\eta$ values indicative of ongoing star-formation and have lower S{\'{e}}rsic indices than conventional red early-types.  The concentration and asymmetry indices are independent of colour, suggesting that the blue ellipticals are smooth systems, in contrast to conventional late-types.  The vertical dashed lines indicate the colour used to define the blue spheroids.  The triangles mark the locations of the example galaxies used in Figure~\ref{fig:de}.}
\label{fig:dE}
\end{figure*}

The galaxies classified as elliptical/lenticular have been reassigned into red and blue classes based on a cut at $u-r=2.0$.  
Panels (a) and (b) of Figure~\ref{fig:dE} show the absolute $B$-band magnitude and surface brightness of the two populations versus rest-frame $u-r$.  The blue galaxies are intrinsically
fainter on average, with only three galaxies having $M_{{\rm B}}<-18$.  This suggests the blue galaxies are less massive than the
red ellipticals.


Variations in galaxy colours can be due to variations in star-formation activity, metallicity, dust or age.  However, it is unlikely that the blue population owes its colour to the presence of dust since this would make the galaxies redder than the conventional ellipticals which are known to possess very little dust.

If the blue galaxies are indeed less massive than the red population, as indicated by their $B$-band luminosities, they will have lower metallicities, since there is a well known mass-metallicity relation for galaxies (see e.g. \citealt{sal05} for a recent analysis).  Thus the mass-metallicty relation is a plausible explanation of the variation in colour, since metal rich galaxies will appear redder.  However, the bimodality cannot be due to 
a mass-metallicity relation for ellipticals/lenticulars only, since the variation in colour is too large (see e.g. \citealt{str01} for the $u-r$ colour distribution for E/S0 galaxies in the SDSS), and the luminosity function of ellipticals is well fitted 
by a Schechter function (e.g. \citealt{efs88}) and thus would not exhibit a bimodal colour distribution for a linear mass-metallicity relation.  Thus, either the blue galaxies are not ellipticals, or there must either be variation in the age or star-formation of ellipticals as a function of mass. 

\citet{mateo98} shows that all dwarf galaxies in the local group have an old stellar population and most also have a young stellar population.  The relative fractions of the two populations and the age of the young stellar population can have a
 strong effect on the galaxy colour.  Hence variation  in star-formation activity could potentially produce a bimodal colour distribution as observed.

The $\eta$ parameter of the 2dF survey is a sensitive indicator of star-formation activity (\citealt{mad02}).
The two classes  show a strong variation in  $\eta$,  panel (c) of Figure~\ref{fig:dE}.  The red
galaxies occupy a narrow range of $\eta$ with values typical of massive early type galaxies (\citealt{mad03}), whereas the blue galaxies have
correspondingly larger values of $\eta$ and a larger scatter.  Thus the red galaxies are mainly quiescent, but the blue galaxies display a range of star-formation activity.  The correlation between colour and star-formation activity is to be expected since galaxy colour and spectra are intimately related, star-forming galaxies have stronger
emission lines and display bluer colours than more quiescent galaxies (\citealt{ken98}).


Panel (d) of Figure~\ref{fig:dE} shows the S{\'{e}}rsic index of the
galaxies when fitted with a S{\'{e}}risc profile only
(\citealt{gra05}).  The blue galaxies have values consistent with pure
exponential systems, whereas the red galaxies have values in keeping
with early-type galaxies.


Although the blue galaxies have many properties in common with late type galaxies they have similar concentration and asymmetry indices to normal ellipticals as shown in panels (e) and (f) of Figure~\ref{fig:dE}, suggesting that they really are smooth systems, and not simply misclassified late-types (a comparison of $C$ with typical late types can be seen in panel (f) of Figure~\ref{fig:bigfig}).

\begin{figure*}
\vspace{-1.0cm}
\centering \includegraphics[scale=0.85,angle=0]{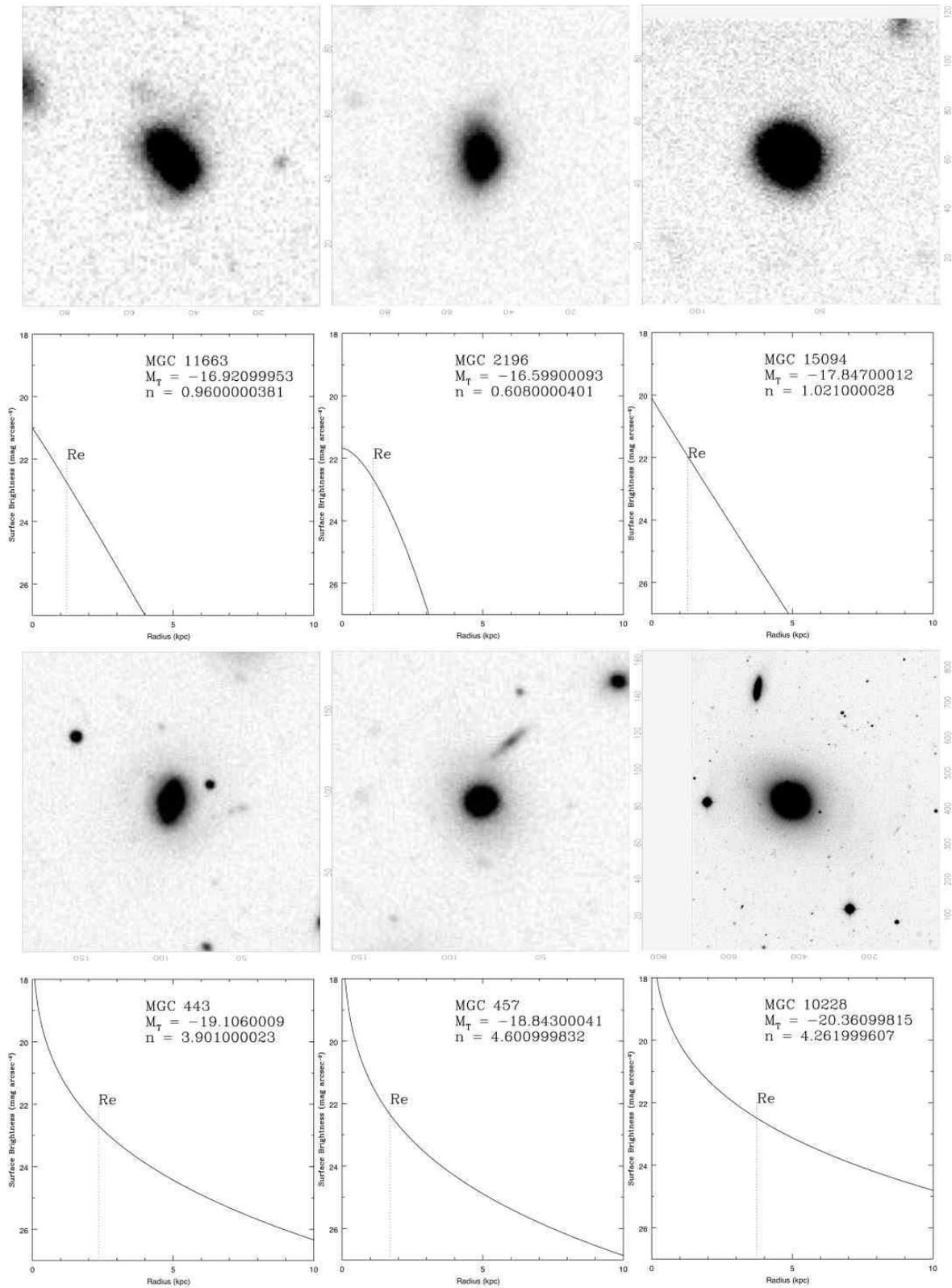}
\vspace{-2.0cm}    
\caption{Examples of galaxies classified as E/S0 by eye but having much bluer colours than conventional ellipticals accompanied by their surface brightness profiles (top row).  The bottom row shows conventional red E/S0 galaxies for comparison.}
\label{fig:de}
\end{figure*}

It seems that the blue galaxies are fundamentally different to both
conventional ellipticals, spirals and late-types.  Their low $B$-band
luminosities suggests they are low mass systems.  They have blue
colours, most likely due to increased star-formation activity, as
evidenced by high values of $\eta$, in common with local dwarf
galaxies (\citealt{mateo98}).  We cannot rule out that they are
misclassified late-type galaxies, e.g.\ with very low surface
brightness discs, but their appearance and concentration suggests that
they are smooth systems.  Some examples of the appearance of these
galaxies are shown in Figure~\ref{fig:de}, along with comparison
images of conventional red E/S0s. Hereafter these galaxies are
referred to as blue spheroids.

\section{Correlations with galaxy morphology}
\label{sec:correl}

The distribution of individual parameters as a function of
morphological type has been examined.  This provides an immediate
understanding of the data set, useful for interpreting the more
complicated statistical analyses, and as an indication of which
parameters are likely to form the most useful input into clustering,
principal components and discrimination analyses.  The distributions
of each parameter are summarised below, and examples are shown in
Figure~\ref{fig:bigfig}.  Statistics of the distributions of each
parameter as a function of galaxy type are given in
Table~\ref{tab:correl}.

\subsection{$B$ band absolute magnitude}

The distribution of $B$ band absolute magnitudes, $M_{{\rm B}}$, are
shown in panel (a) of Figure~\ref{fig:bigfig}.  The distributions for
all classes are similar with the exception of the blue spheroids,
which are preferentially faint, indicating that they are less
massive systems.  There is a mild trend for galaxies to become fainter
going from E/S0 to Sabc to Sd/Irr to blue spheroid, but it is less
than 68\% significant.


\subsection{$B$ band surface brightness}

The distribution of $B$ band surface brightness, $\mu_{{\rm B}}$, is shown in panel (b) of Figure~\ref{fig:bigfig}.  The E/S0 galaxies are generally bright, whilst the Sabc and Sd/Irr are $\sim 1.5$ mag arcsec$^{-2}$ fainter, but with a large scatter in the surface brightness of all types.  The blue spheroid galaxies have a large spread in values with a peak intermediate to E/S0 and Sabcd/Irrs.

\subsection{Colour}

Galaxy colours have been computed using $u-r$ as \citet{str01} have shown that this combination can provide nearly optimal separation into two galaxy
types for SDSS.  Histograms of the colour distribution are shown in panel (c) of Figure~\ref{fig:bigfig}. 


Recall that the distinction between E/S0 and blue spheroid is artificial since blue spheroid galaxies are defined as having colours $u-r \le 2.0$.  There is a trend for
galaxies to become bluer going from E/S0 to Sabc to Sd/Irr to blue spheroid, although there is a large overlap between any two adjacent classes.

\subsection{2dFGRS $\eta$ parameter}

Panel (d) of Figure~\ref{fig:bigfig} shows $\eta$ as a function of type.  There is a rather clear demarcation between E/S0 galaxies and all the later types, and thus $\eta$ shows some promise
 of a morphological classifier.  Indeed \citet{mad03} discusses in detail the ability of $\eta$  to distinguish between early (E/S0) and late (spirals and irregulars) type galaxies.  In that paper it is found that the commonly used cut of $\eta=-1.4$ compares very favourably with morphological classifications based on more complicated
 combinations of the first nine principal components of the spectra ($\eta$ is a linear combination of the first two).



The ability of $\eta$ to distinguish between galaxies of early and late morphologies is reasonably accomplished, however, further discrimination of different types is less robust.  Panel (d) of Figure \ref{fig:bigfig} shows a large degree of overlap between Sabcs,
 Sd/Irrs and blue spheroids.  
 This will be shown to be a general feature throughout our attempts to provide a system 
of galaxy classification.  It is relatively easy to separate E/S0 galaxies from Sabcd/Irrs but
 further distinction is extremely problematic.  The comparatively large spread in $\eta$ of the spiral galaxies is also seen in the morphologically selected sample of \citet{mad03}, and in an analysis by \citet{nor02} based on APM galaxies.

\citet{mad03b} show that $\eta$ is much more tightly correlated with the rate of current star-formation in a galaxy (which itself may be correlated with morphology).  
\citet{mad02} interpret $\eta$ as a measure of the current
star-formation rate in the galaxy, but they also show that $\eta$ is
correlated with morphology for a sample of 21 galaxies from the
Kennicutt Atlas (\citealt{ken92}).  This sample classifies galaxies
individually into subtypes of E/S0, Sa, Sb and Scd.  It is seen that
$\eta$ is progressively larger for later spiral types.  Thus it could
be argued that the large spread in our Sabc types seen in
Figure~\ref{fig:bigfig} is a result of binning the sub-types Sabc into
a single bin.  However the rather low $\eta$ values of some Sd/Irrs
and the overlap between the Sabcs and E/S0s, suggest that there is
some intrinsic spread in the values of $\eta$ with sub-types (this
may in part be exacerbated by fibre-placement issues.  For example, it is possible that rather different $\eta$ values would result from a fibre being placed on the bulge or the disc of a spiral galaxy).



\subsection{S{\'{e}}rsic index}

The S{\'{e}}rsic indices, $n$, of Sabc, Sd/Irr and blue spheroid
galaxies fitted with S{\'{e}}rsic profiles only (no exponential disc,
\citealt{gra05}) are distributed around 1, indicating an exponential
surface brightness profile.  On the other hand E/S0 galaxies have
higher values of S{\'{e}}rsic index scattered around the de
Vaucouleurs value of $n=4$.  Panel (e) of Figure~\ref{fig:bigfig}
shows the histograms.


\subsection{Concentration and Asymmetry Indices}

\begin{figure*}
    \centering \includegraphics[scale=0.8,angle=0]
    {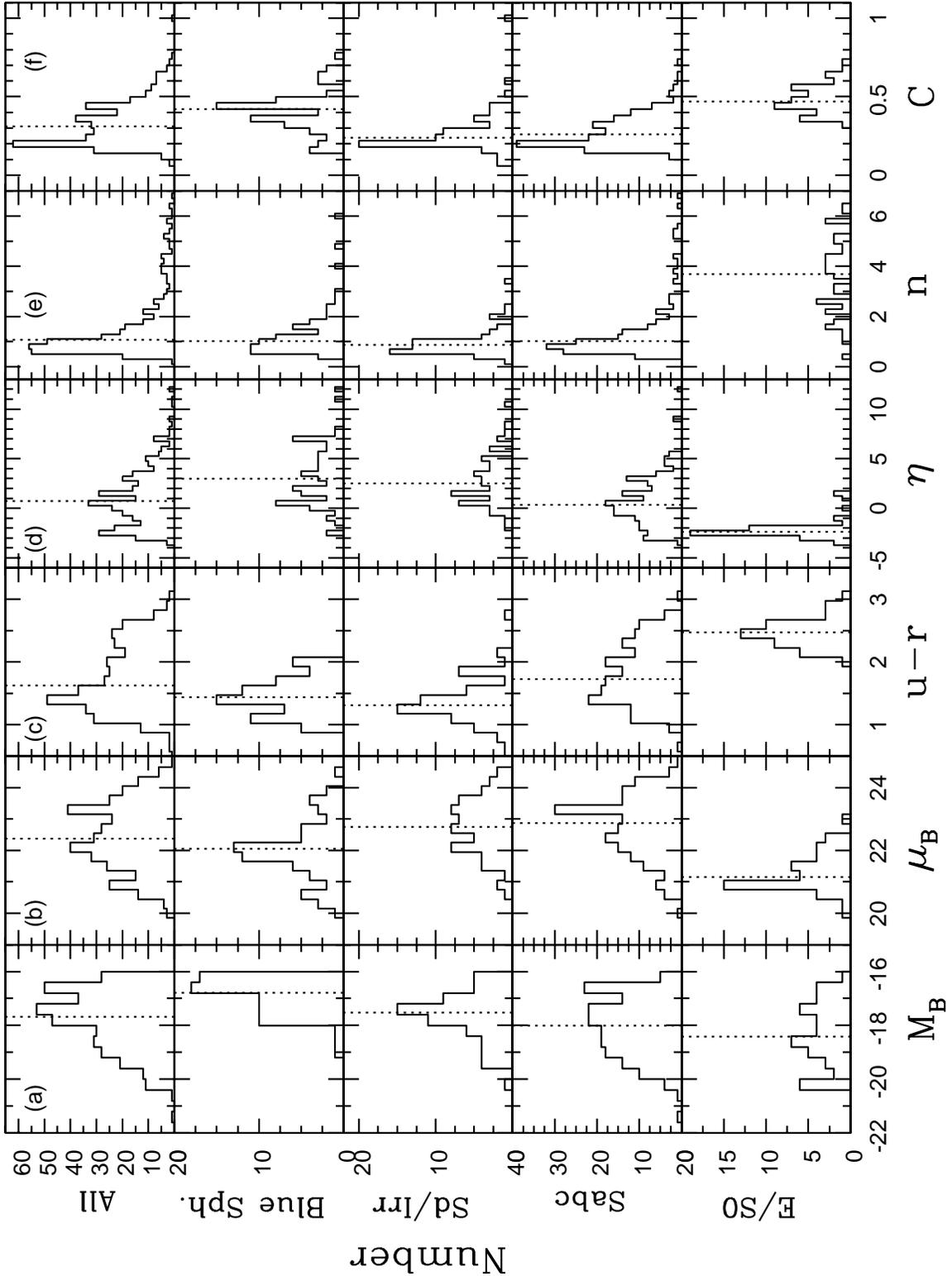}
    \caption{Histograms of various parameters as a function of galaxy type.  The median of each distribution is shown by the a dotted line.  Panel (a) shows the distribution of $M_{{\rm B}}$; (b) shows the distribution of $\mu_{{\rm B}}$; (c) shows the distribution of $u-r$; (d) shows the distribution of $\eta$; (e) shows the distribution of Sersic indices; (f) shows 
the distribution of concentration indices.}
\label{fig:bigfig}
\end{figure*}

The concentration indices, $C$, are shown in panel (f) of
Figure~\ref{fig:bigfig}, defined as 
the fraction of light within 10\% of the normalised radius (see \citealt{sim02}).
The distribution of Sabcs is very similar to the
distribution of Sd/Irrs showing an unconcentrated distribution, while
the the distribution of blue spheroids is very similar to the
distribution E/S0s and has higher concentration.  This suggests that
blue spheroids are not simply misclassified Sabcd/Irrs, and provides
support for the original visual classifications.


The asymmetry indices, $A$, are similar for all classes of galaxies except Sd/Irrs which are slightly more asymmetric.

\begin{table}
\caption{Statistics of the distributions of key parameters as a function of galaxy type.  An idea of skewness may be estimated by comparing the
mean and standard deviation with the biweight estimators.  For Gaussian distributions the values from each should be identical.  In a heavily
skewed distribution the biweight estimators are probably more useful.}
\label{tab:correl}
\begin{tabular}{lccccc}
&Mean&$\sigma$ & $c_{{\rm bi}}$ & $s_{{\rm bi}}$ & Median \\ \hline
&\multicolumn{5}{c}{ $\mathbf{M_{{\rm B}}}$} \\
E/S0 & -18.32 &1.13 &-18.32 &1.18 &-18.41\\
Sabc & -18.07 & 1.09 & -18.04 & 1.12 & -18.00\\
Sd/Irr & -17.64 & 0.89 & -17.57 & 0.91 & -17.53\\
Blue spheroid & -16.95 & 0.67 & -16.87 & 0.67 & -16.78\\
E/S0 \& blue spheroid & -17.50& 1.11& -17.32& 1.10& -17.25\\
All & -17.81 & 1.10 & -17.74 & 1.12 & -17.68\\
&\multicolumn{5}{c}{ $\mathbf{\mu_{{\rm B}}}$} \\
E/S0 & 21.30 & 0.61 & 21.22 & 0.62 & 21.16\\
Sabc & 22.77 & 0.97 & 22.82 & 1.01 & 22.87\\
Sd/Irr & 22.77 & 0.90 & 22.78 & 0.93 & 22.75\\
Blue spheroid & 22.01 & 0.98 & 21.99 & 1.01 & 22.05\\
E/S0 \& blue spheroid & 21.72 & 0.92 & 21.66 & 0.92& 21.67\\
All & 22.42 & 1.07 & 22.42 & 1.11 & 22.38\\
&\multicolumn{5}{c}{ $\mathbf{u-r}$} \\
E/S0 &2.46& 0.22 & 2.46 & 0.22 & 2.47 \\
Sabc & 1.79 & 0.52 & 1.78 & 0.52 & 1.73\\
Sd/Irr &1.39 & 0.57 & 1.32 & 0.36 & 1.31\\
Blue spheroid & 1.44 & 0.28 & 1.43& 0.29 & 1.44 \\
E/S0 \& blue spheroid & 1.85 & 0.56 & 1.83 & 0.60 & 1.71\\
All & 1.74 & 0.57 & 1.70 & 0.57&  1.63 \\
&\multicolumn{5}{c}{ $\mathbf{\eta}$} \\
E/S0 &-2.10 &  1.05 & -2.40 & 0.54 & -2.36 \\
Sabc & 0.71 &  2.47 & 0.45 & 2.30 & 0.36 \\
Sd/Irr &3.26 &  3.92 &  2.51 &  2.65 & 2.50\\
Blue spheroid & 3.34 & 3.50 & 2.97 & 3.13 & 2.96 \\
E/S0 \& blue spheroid &1.14 & 3.86 & 0.69 & 3.63 &  0.45\\
All & 1.32 & 3.40 & 0.88 & 2.93 & 0.74 \\
&\multicolumn{5}{c}{ \bf{S{\'{e}}rsic Index}} \\
E/S0 &3.49 & 1.48 & 3.52 & 1.55 & 3.68\\
Sabc & 1.45 & 1.22 & 1.02 & 0.60 & 1.02\\
Sd/Irr &1.43 & 3.64 & 0.83 & 0.38 & 0.87 \\
Blue spheroid & 1.37 & 0.98 & 1.09 & 0.62 & 1.02 \\
E/S0 \& blue spheroid & 2.22 & 1.60 & 1.80 & 1.60 & 1.63\\
All & 1.70 & 2.03 & 1.07 & 0.74 & 1.07 \\
&\multicolumn{5}{c}{ \bf{ Concentration index (inner 10\%)} } \\
E/S0 &0.48 & 0.09 & 0.48 & 0.10 & 0.47 \\
Sabc &0.28 & 0.11 & 0.26 & 0.10 & 0.26 \\
Sd/Irr & 0.27 & 0.13 & 0.24 & 0.09 & 0.24\\
Blue spheroid & 0.41 &  0.13 & 0.40 &  0.13 & 0.42\\
E/S0 \& blue spheroid & 0.44 & 0.12 &  0.44 & 0.12 & 0.44\\
All & 0.33 & 0.14 & 0.32 & 0.14 & 0.31 \\
&\multicolumn{5}{c}{ \bf{ Asymmetry index (inner 10\%)}} \\
E/S0 & 0.06 & 0.03 & 0.05 & 0.03 & 0.05\\
Sabc & 0.08 & 0.06 & 0.07 & 0.03 & 0.07\\
Sd/Irr & 0.13 & 0.08 & 0.11 & 0.06 & 0.11\\
Blue spheroid & 0.06 &  0.04 & 0.06 & 0.03 & 0.06 \\
E/S0 \& blue spheroid & 0.06 & 0.04 & 0.06 & 0.03 & 0.05\\
All & 0.08 & 0.07 &  0.07 & 0.04 & 0.07
\end{tabular}
\end{table}


\section{Principal components analysis}
\label{sec:pca}

The parameter set was examined in more detail by considering groupings and correlations between several parameters simultaneously. 
Principal components analysis (PCA, \citealt{mur87}) identifies combinations of parameters which summarise the distribution of the whole set.  The principal components are the projections onto the eigenvectors which successively contain the maximum variance, under the
condition that all components are mutually orthogonal.  This can be thought of as finding a natural set of axes throughout the parameter space
which minimise the spread of the parameters.  That is, if the parameters are pictured as a cloud of points in space, PCA finds the axes
to which the points are closest, where closeness is defined as the Euclidean distance between vectors.

Thus PCA can be used to reduce the dimensionality of the parameter space by projecting the data onto the first few axes. 
Important parameters can be identified by examining the components of the eigenvectors.

\subsection{Method}

A good introduction to PCA is given by \citet{mur87}.  The method is summarised here. 

 PCA takes as its input a selection of parameters from the whole catalogue.  This forms
 an $N \times M$ matrix, ${\bf A}$, where $M$ is the number of parameters and $N$ is the number of galaxies.  A new matrix, ${\bf B}$, is
 formed by subtracting the mean values of each parameter from every element in the matrix, and normalising by the standard deviation i.e., 

\begin{eqnarray}
\label{eqn:matrixb}
b_{ij}=\frac{a_{ij}-\bar{a}_{j}}{\sqrt{\frac{\sum_{i=1}^{N}\left( a_{ij}-
\bar{a}_{j} \right) ^{2}}{N}}}, & 1 \le i \le N, & 1 \le j \le M 
\end{eqnarray}

where $b_{ij}$ are the elements of ${\bf B}$ and $a_{ij}$ are the elements of ${\bf A}$ and

\begin{equation}
\bar{a}_{j}= \frac{1}{N}\sum_{i=1}^{N}a_{ij}.
\end{equation}   
The covariance matrix of ${\bf B}$ is then formed,

\begin{eqnarray}
c_{jk} = \frac{1}{N}\sum_{i=1}^{N}b_{ij}b_{ik}, & 1 \le j \le M , & 1 \le k \le M 
\end{eqnarray}

where $c_{jk}$ are the elements of the covariance matrix ${\bf C}$.  Matrix ${\bf C}$ is real and symmetric, and therefore has real, positive
eigenvalues.  The numerical order of these eigenvalues yields the strength of the variance of the data in the corresponding eigenvector.  

The principal components, ${\bf P}$, can then be found by projecting the eigenvectors onto the normalised, mean-subtracted matrix ${\bf B}$, i.e.\,

\begin{equation}
{\bf P}={\bf BC}.
\end{equation}

The order of the principal components is found from the numerical order of their corresponding eigenvalues, so $pc1$ are the elements of the $N \times 1$ column vector corresponding to the highest eigenvalue, etc.

\subsection{Results}

A PCA was carried out on various combinations of the parameters $\{M_{{\rm B}},\mu_{{\rm B}},u-r,\eta,n,C,A\}$ in order to determine any particular combinations of parameters which account for a large amount of
variance, and hence provide a useful description of galaxy properties.  

Many combinations of parameters were experimented with.  In no experiment did any individual parameter dominate the principal eigenvectors: no single parameter had significantly more variance than any other.  Considering the parameters as a
cloud of points, they are not elongated in any way which is aligned to an individual parameter.   Compression of the data required several parameters
to be combined together, indicating that photometric ($M_{{\rm B}}$, $\mu_{{\rm B}}$, $u-r$), spectroscopic ($\eta$) and structural ($n$, $C$, $A$) parameters are all of similar importance in describing the distribution of galaxies throughout the parameter space.

\begin{figure}
\begin{minipage}[c]{0.3\textwidth}
    \centering \includegraphics[scale=0.4,angle=0]
    {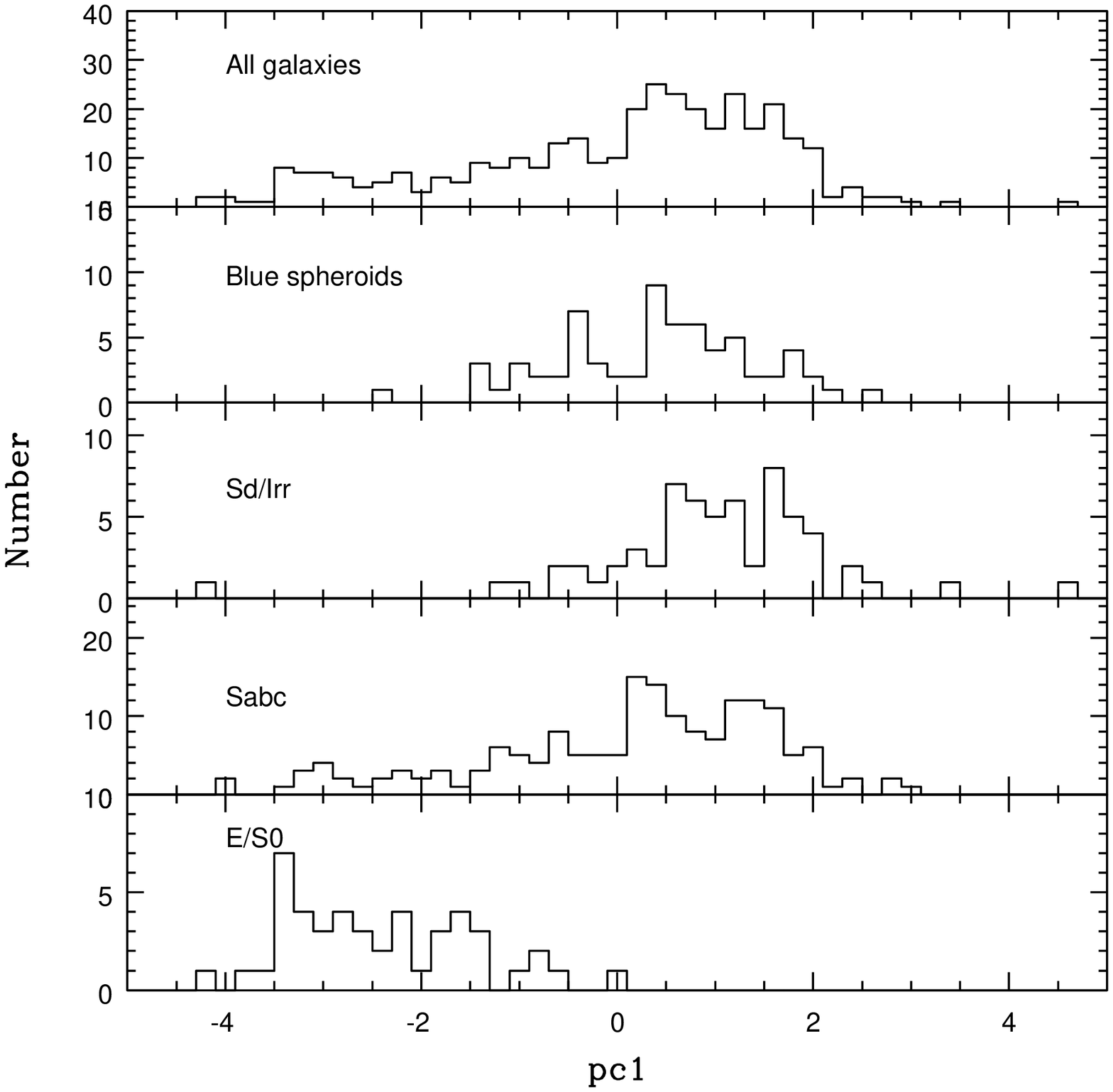}
 \end{minipage}
 \begin{minipage}[c]{0.3\textwidth}
    \centering \includegraphics[scale=0.4,angle=0]
    {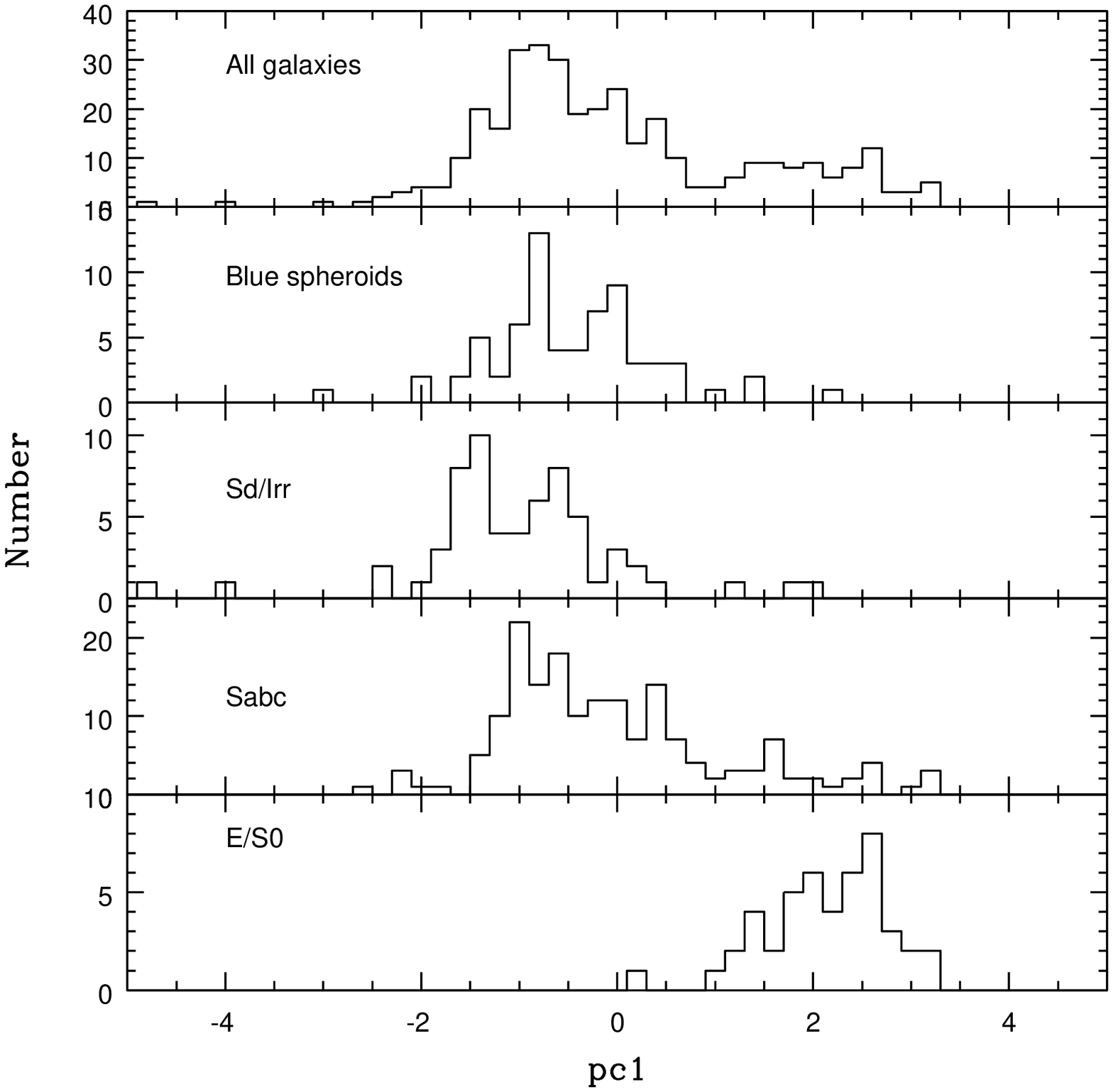}
 \end{minipage}
 \caption{The distribution of the first principal component, $pc1$, for galaxies of different type.  The top panel shows the results of PCA using the set $\{M_{{\rm B}},\mu_{{\rm B}},u-r,\eta,n,C,A\}$, and the bottom panel using the set $\{u-r,\eta,n,C\}$
.  In both cases the first principal component clearly separates early and late type galaxies.}
\label{fig:pca_type}
\end{figure}

The first principal component usually projected the galaxies in such a way that the E/S0 galaxies were separated from the other types of galaxies.  Thus, the parameter space has a maximum variance in a direction that separates E/S0 galaxies from other types.  Note that PCA is not given any information regarding the visual morphological classifications.  The distinction between E/S0 and other galaxies is clearly of fundamental importance in terms of their observational properties; influencing photometric, spectroscopic and structural parameters.  Two examples of this are shown in Figure~\ref{fig:pca_type} using the subsets $\{M_{{\rm B}},\mu_{{\rm B}},u-r,\eta,n,C,A\}$ and $\{u-r,\eta,n,C\}$, illustrating the distinction between E/S0 and the other galaxies.  Note that the distribution of Sabcs is skewed with a tail overlapping with the E/S0s.  Whilst
 we cannot rule out misclassification for all the Sabc galaxies in the tail, some of them display very clear spiral arms.  The other principal components showed no clear distinction between galaxies of different types.  No combination of principal components was found which could separate Sabc, Sd/Irr and blue spheroid galaxies.

A caveat is to note that information on the errors associated with the parameters was not included in the analysis, since errors are not known for all parameters, e.g.\ $\eta$.  If some parameters have
large errors associated with them they may skew the results.

\section{Linear discriminant analysis}
\label{sec:lda}

Linear discriminant analysis (LDA) defines combinations of parameters which optimally discriminate between objects of known classes.  We have used LDA 
with various combinations of parameters to discriminate between galaxies of different morphology.  Thus we can assess the relative merit of the input parameters in assigning, and hence correlation with, galaxy morphology.


\subsection{Method}


Consider again the $N \times M$ matrix ${\bf B}$ of equation~\ref{eqn:matrixb}, containing the normalised, mean subtracted properties of a sample of galaxies, where $N$ is the number of galaxies and $M$ is the number of measured properties.   Assuming that the sample has been divided into two classes, 0 and 1, the vectors, ${\bf m_{0}}$ and ${\bf m_{1}}$, describing the means of each property, and the covariance matrices, ${\bf V_{0}}$ and ${\bf V_{1}}$, of each class may be calculated.





The combined covariance matrix, ${\bf W}$, of each class may then be computed, 

\begin{equation}
{\bf W}={\bf V}_{0}+{\bf V}_{1}.
\end{equation}

Fisher's linear discriminant is defined such as to maximise the distance between the samples means after standardising by the variance within the samples (\citealt{han81}, page 83).  That is, we want to maximise the ratio,

\begin{equation}
\label{eqn:fisher}
\frac{{\bf m_{0}}-{\bf m_{1}}}{{\bf Wf}},
\end{equation}

where ${\bf f}$ is a vector describing the set of weights which maximise equation~\ref{eqn:fisher}.  Thus,

\begin{equation}
{\bf f} \propto {\bf W}^{-1}({\bf m_{0}}-{\bf m_{1}}).
\end{equation}

Finally we may define a new vector containing the initial input data weighted so as to maximise the difference between class 0 and 1,

\begin{equation}
{\bf t}=({\bf m_{{\rm 0}}}-{\bf m_{{\rm 1}}})^{{\rm T}}{\bf W}^{-1}{\bf B}.
\end{equation}

Thus Fisher's LDA must first be applied to a population of galaxies whose types are known (in our case this is the whole catalogue).  Thereafter an appropriate division can be made
based on a value of $t$ (where $t_{i}$ are the elements of vector ${\bf t}$).  The choice of $t$ will depend on how
`pure' a sample is desired, setting $t$ to give little contamination in one class, will give a larger contamination in the other class if
the two classes overlap. 

\subsection{Results}

The data were first standardised to zero mean and unit variance before performing the analysis.  We experimented with various combinations of parameters.  It was relatively easy to separate E/S0s and the rest.  Discriminating between other types of galaxies was always less successful.  Some examples are given below.

\begin{figure}
\centering \includegraphics[scale=0.4,angle=0]{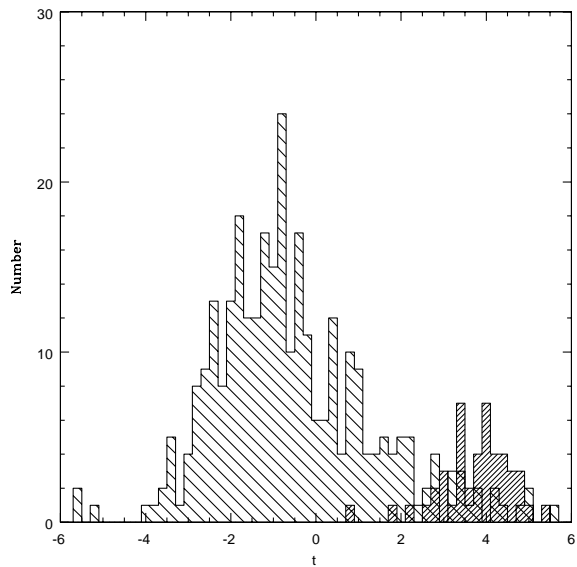}
\caption{Example of linear discriminant analysis.  The closely hashed histogram on the right hand side represents E/S0 galaxies and the open hashed histogram the rest.  A cut at
$t(x)=2.40$ correctly classifies 92\% of E/S0 galaxies.}
\label{fig:lda}
\end{figure}

The input set $\{M_{{\rm B}},\mu_{{\rm B}},u-r,\eta,n,C,A\}$
 was used.  The resulting discrimination is illustrated in Figure~\ref{fig:lda}.  A cut at $t(x)=2.40$ resulted in 93\% of E/S0s being
correctly assigned and 93\% of others.  This same cut correctly classified 99\% of galaxies below the cut and 66\% above.  Thus, the separation of E/S0 from the rest of the galaxies was relatively successful cf.\ 80\% agreement when classifying by eye.

In contrast Sabcs could only be separated with $72\%$ success, Sd/Irrs with $75\%$ success and blue spheroids with $81\%$ success. 

An input set consisting of purely structural parameters $\{n,C,A\}$
was experimented with.  This was less successful than the full parameter set, but  E/S0s could still be separated with 83\%
success.

Similarly star-forming parameters $\{ \eta$, $u-r \}$ were investigated.  These two parameters alone could separate E/S0 galaxies with 88\% success,
but had poorer success rates at separating any other types of galaxies.


\section{Clustering Analysis}
\label{sec:hca}

So far the work presented has concentrated mainly on correlations with the morphological types assigned by eye, and combinations of
parameters to optimise discrimination between these \emph{a priori} defined classes.  It may be the case however, that there are natural
classes of objects grouped within the parameter space which do not necessarily reflect morphological types.
A hierarchical clustering analysis (HCA) has been performed in order to search for such natural groupings within the parameter set.  The method and
the results are described below.

\subsection{Method}

$N$ galaxies, each described by $M$ parameters, are described by matrix $\bf{A}$, as in PCA.  Each galaxy can be thought of as a point in a
parameter space.  The \emph{distance} between each point and all the others is calculated yielding a set of $N(N-1)/2$ distances, where the
distance will be defined below.  Once the distances are known clusters are made by agglomerating points within the parameter space based on
their distances.  For example, the smallest distance $d_{ik}$ is determined, and objects $i$ and $k$ are then agglomerated and replaced with a new
object $i \cup k$.  The set of distances are then updated using the new object in place of $i$ and $k$.

Several methods can be used to define distances, and the centres of the newly created clusters.  The method used here is that of
\citet{war63}, which defines the distance between two clusters as the amount by which the total distance from each point to the centre of its cluster would increase by merging the two clusters (technically, it is the sum of the squares of the Euclidean distances from the objects to the joint cluster mean minus the sum of squares from the
objects to their individual cluster means, so the distance is the amount the sum of squares would increase if the clusters were to be merged, see \citealt{han81}).
This method was implemented using freely available {\sc java} software written by
F.~Murtagh\footnote{http://astro.u-strasbg.fr/$\sim$fmurtagh/mda-sw/}. 

\subsection{Results}

Again experiments were performed with the input set $\{M_{{\rm B}},\mu_{{\rm B}},u-r,\eta,n,C,A\}$, and parameters were standardised on input.  Galaxies were split into only two clearly distinct classes.

\begin{figure}
\begin{minipage}[c]{0.5\textwidth}
\centering \includegraphics[scale=0.3,angle=270]{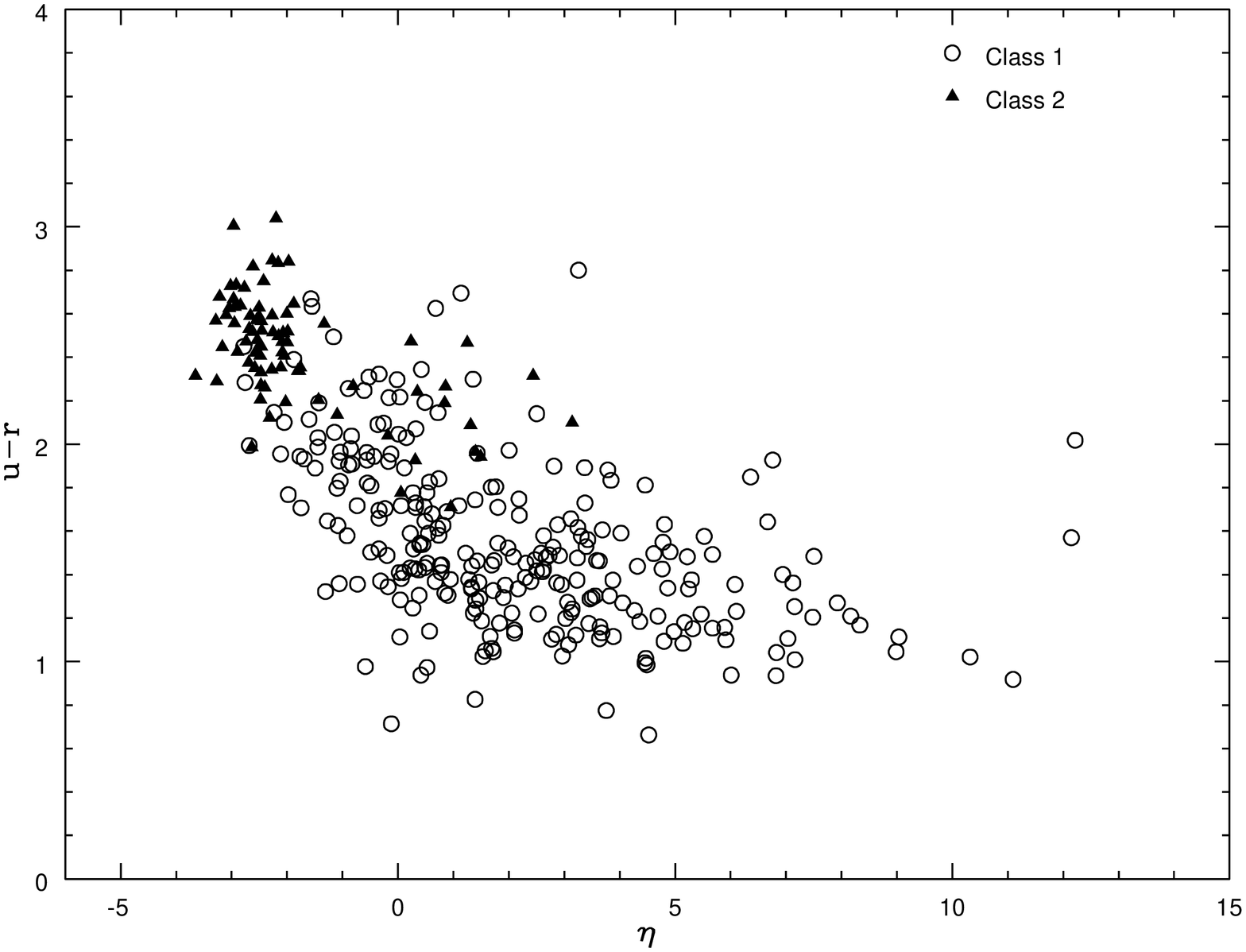}
\end{minipage}
\begin{minipage}[c]{0.5\textwidth}
\centering \includegraphics[scale=0.3,angle=270]{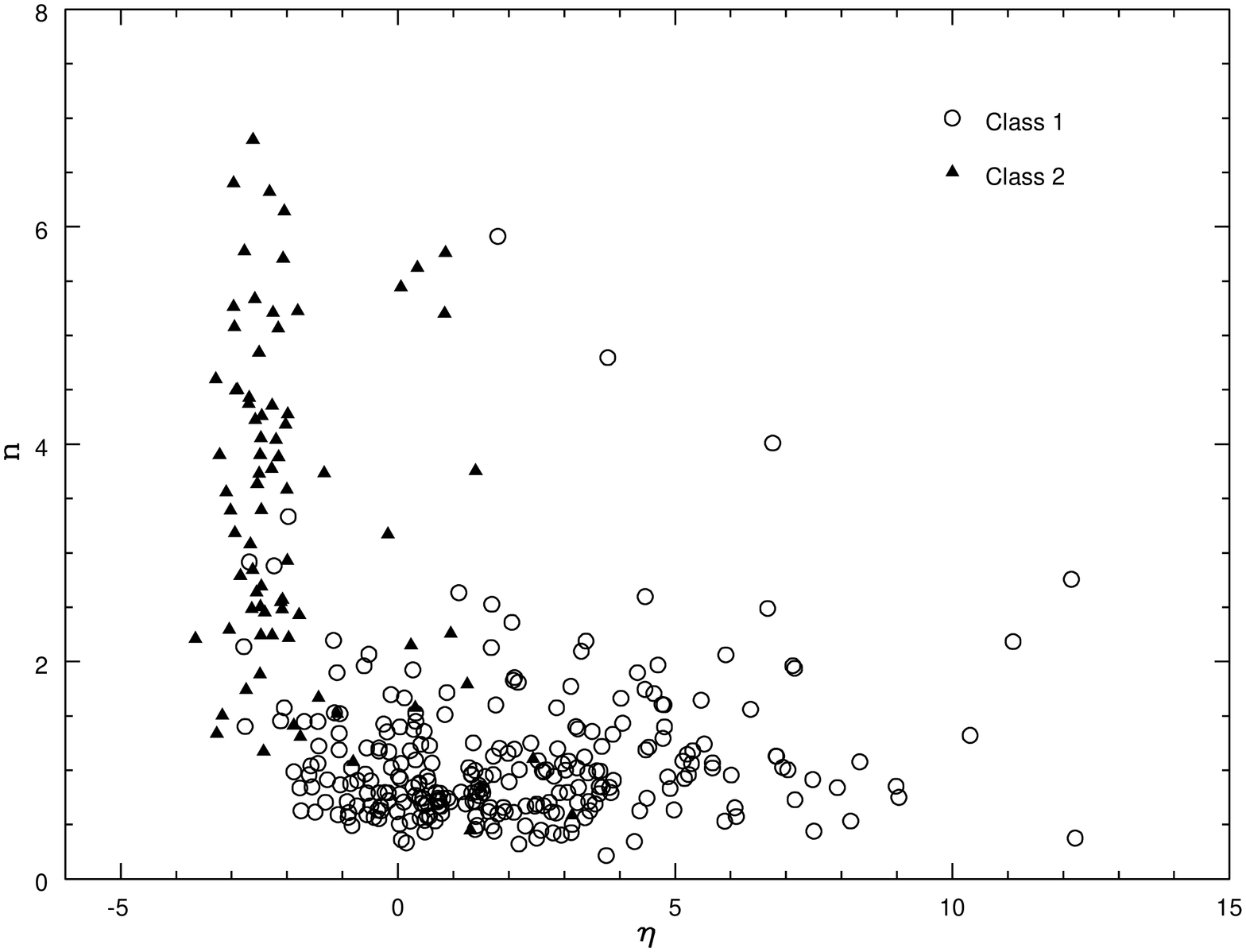}
\end{minipage}
\begin{minipage}[c]{0.5\textwidth}
\centering \includegraphics[scale=0.3,angle=270]{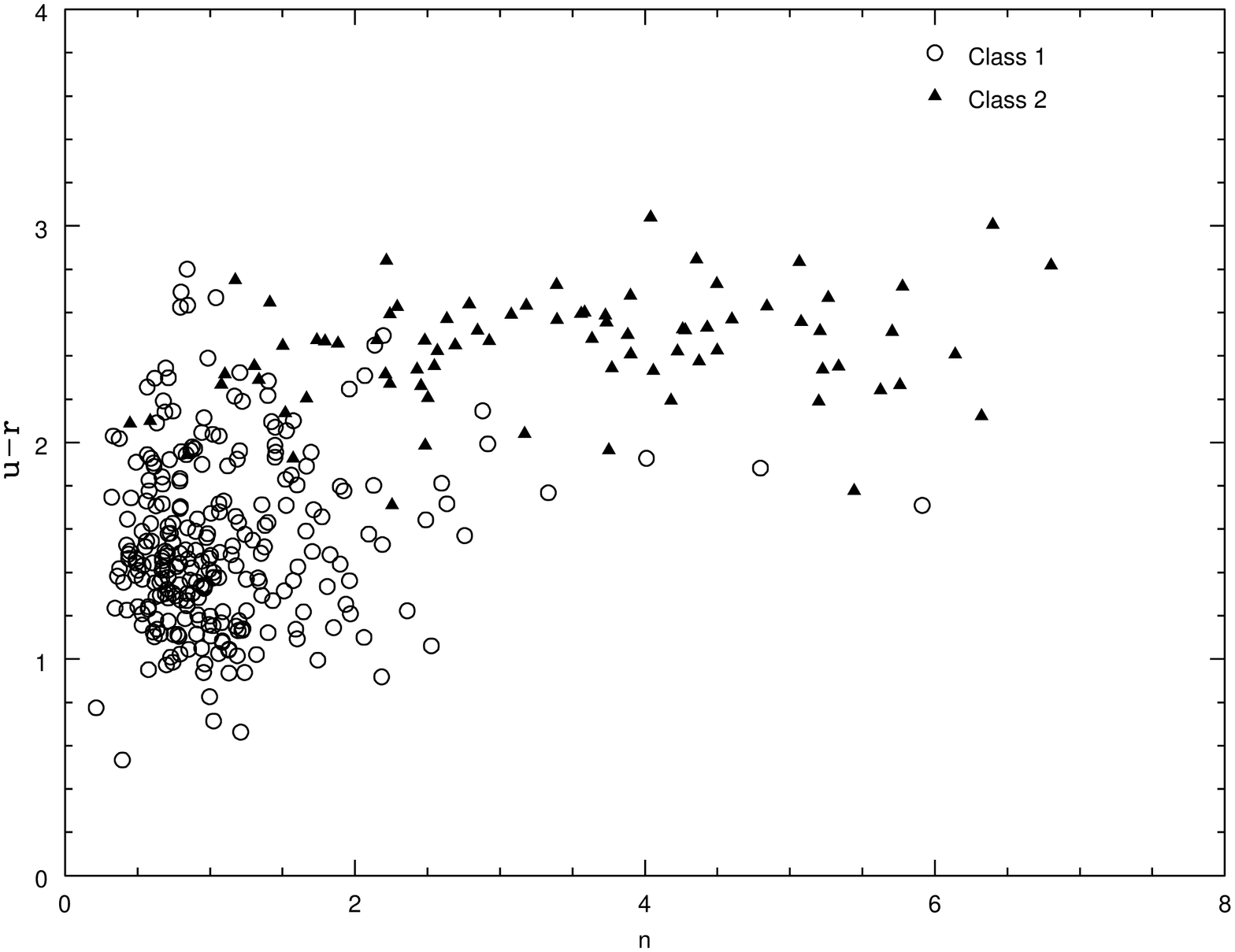}
\end{minipage}
\caption{Galaxies classified into natural categories as a function of rest-frame $u-r$, S{\'{e}}rsic index and $\eta$.}
\label{fig:clus}
\end{figure}

Classifying galaxies into two categories which gave the greatest separation, reveals a segregation of galaxies strongly dependent on $u-r$,
$\eta$ and the Sersic index, $n$.  Classification into more than two categories did not reveal any strong trends with any parameter for the third
class of object.
Figure~\ref{fig:clus} shows the dependence of the two natural classes on $u-r$, $\eta$ and $n$.

\begin{figure}
\centering \includegraphics[scale=0.4,angle=0]{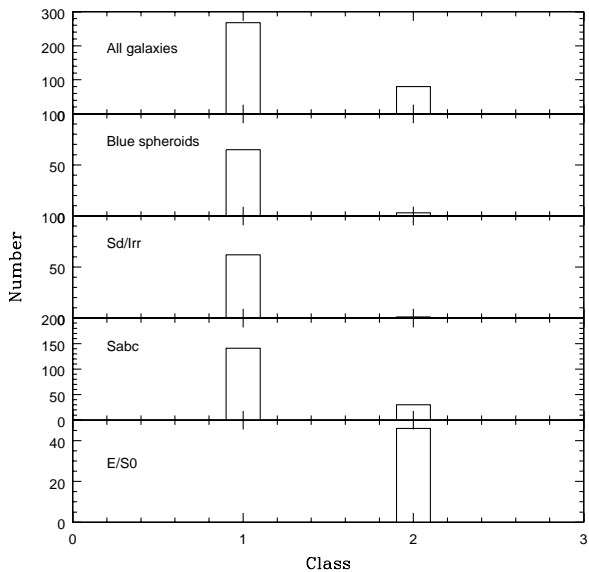}
\caption{The distribution of galaxies within each cluster per morphological type.}
\label{fig:clus_type}
\end{figure}

The clustering of the galaxies in these natural classes was tested for correlations with morphology.  Membership of group 1 or 2 was
assigned as an extra input parameter into a LDA.  No improvement (or deterioration) was made over previous efforts to discriminate galaxies into known
morphological types.  This indicates that the classes defined from hierarchical clustering are no more closely correlated with morphology than the
input parameters, although
this could be due to poor classification of galaxies when assigning types by eye.  Figure~\ref{fig:clus_type} shows the distribution of each
morphological type within the clusters.  One cluster is made almost solely of E/S0 galaxies with the majority of
all other types occupying the other cluster.  Although it is possible that some of the galaxies in class 2 of Figure~\ref{fig:clus_type} were originally classified as Sabcd/Irr/blue spheroid as a result of misclassification, some of them have definite spiral morphology.

Using the same input parameters LDA could reproduce the
classifications of the hierarchical clustering analysis with excellent success of 95\% (with 98\% of galaxies below the cut, and 84\% above correctly classified), showing that there is little overlap between the two
classes in this parameter space.

\section{Discussion}
\label{sec:discuss}

We have examined the distributions of photometric ($M_{{\rm B}}$, $\mu_{{\rm B}}$, $u-r$), spectral ($\eta$) and structural ($n$, $C$, $A$) parameters from a well defined sample of nearby galaxies culled from the Millennium Galaxy Catalogue.  Galaxies were classified into broad morphological types by visual inspection.  

\subsection{Blue spheroids}

A significant fraction of galaxies classified as ellipticals were
found to have some properties (colour, $\eta$ and $n$) more in keeping
with late-type galaxies, but with concentration and asymmetry indices
more typical of early-type galaxies.  These galaxies are
preferentially faint, suggesting that they are low mass galaxies.   
A
significant fraction of these galaxies appear to be undergoing
significant star-formation.  Such blue spheroids have previously been
reported elsewhere, e.g.\ in studies of faint field galaxies in HST
images (\citealt{dri95}). This population of blue spheroids may well
be responsible for the highly discrepant ``early-type'' luminosity
function estimates (see \citealt{delap03} for discussion). Early-types
cuts by colour or spectral type will exclude this population whereas
morphological cuts will not.  Note also that the high fraction of blue spheroids in our sample is probably a result of the near volume-limited nature of the sample, the fraction would likely be much lower in a magnitude limited sample.

According to hierarchical theories of structure formation, less
massive galaxies form at earlier epochs than massive galaxies.  This
leads to the conclusion that the stellar populations of massive
galaxies should be younger than those of dwarf galaxies, unless 
star-formation is not coincident with the epoch of galaxy assembly.
Observationally, however, there is a lot of evidence that less massive
galaxies undergo star-formation at later epochs than massive galaxies.
Local dwarf galaxies are seen to be composed of significant fractions
of young stars, perhaps predominantly so (\citealt{mateo98}).  Less
massive galaxies are more likely to have undergone recent star-bursts
than massive galaxies (\citealt{cow96}; \citealt{gav96};
\citealt{kau03}).  The colour-magnitude relation of galaxies in groups
at high redshift shows a truncation at the faint end, beyond which the
typical red galaxies are supplanted by a blue population
(\citealt{kod04}; \citealt{del04}).  Similarly, \citet{dep03b} shows
that the increased fraction of blue galaxies in clusters at high
redshift (e.g.\ \citealt{but84}) may be driven by low mass galaxies
which have recently undergone star-formation.

The blue spheroid galaxies found in the present study are consistent
with this picture of down-sizing in galaxy formation.
The galaxies are centrally concentrated, smooth, and appear as normal
ellipticals, suggesting they have not undergone any recent mergers
which could trigger star-formation, but they possess colours and
$\eta$ values characteristic of star-forming galaxies. Finally it is
worth noting that the statistical tools were unable to isolate the
blue spheroid population from the late types. However the eyeball classifications
did. This implies that there is additional useful information that can
be measured. This might include smoothness (e.g., CAS analysis;
\citealt{con03}) or low frequency Fourier modes (e.g., \citealt{ode02}
for example).

\subsection{Classification and discrimination of galaxy types}

The over-riding feature of all the multi-variate statistical studies was the existence of just two clear groups of galaxies (albeit with some overlap in their properties), corresponding to early types (E/S0) and the rest (Sabcd/Irr/blue spheroid).  The physical difference being the colour, profile shape and the amount of star-formation activity.  Further distinction of separate groups was always less convincing. 

The class of galaxies comprised of Sabcd/Irr/blue spheroids is distributed remarkably smoothly throughout the parameter space, and distinction between any of the morphological types based on their global photometric, spectral and structural properties is difficult and unconvincing.  This is not to say there are no correlations of galaxy properties  with morphology.  For example, \citet{rob94} find correlations (notwithstanding large dispersions) with Hubble type and average values of total mass, mass fraction of neutral hydrogen, total surface density and integrated galaxy colour, amongst others.  These correlations are easily understood in terms of variation in star-formation activity as a function of morphology.  However, the transition of properties from spiral to irregular galaxies is always smooth, in keeping with their occupation of a single class, and results largely from the variation in mass along the sequence.  On the other hand there can be large discontinuities between the properties of elliptical and spiral galaxies, e.g.\ HI mass fraction.

It is plausible then, that the astrophysical processes governing the formation and evolution of galaxies are the same for all morphological types, Sa to Irr, with minor variations due to different masses of the galaxies, e.g.\ the colour differences can be qualitatively explained by the relation between mass and metallicity.  Ellipticals and lenticulars however, appear to require different astrophysical processes.
 
The real and fundamental separation of these two groups needs to be
understood in the context of galaxy evolution.  The existence of two
distinct groups can be explained either in terms of evolution of
galaxies
or in terms of initial conditions.

Perhaps the simplest explanation is that the difference is a result of
 initial conditions, with the two types of galaxies having different
 formation mechanisms.  This requires that ellipticals collapsed early
 in a dissipationless manner, and hence contain old stellar
 populations and a three-dimensional appearance.  Spirals collapsed
 dissipatively, and hence formed into a disc with a reservoir of gas.
 The difficulty in this case lies in explaining what causes galaxies
 to collapse in different ways.  Also this idea is at odds with the
 cold dark matter paradigm, which predicts that galaxies assemble in a
 hierarchical manner, through a series of mergers (\citealt{bau96}).

\citet{hub36} proposed that the morphological sequence of galaxies
represents an evolution from the simple elliptical forms to the grand
design of Sc galaxies.  Much modern research proposes morphological
evolution in the opposite direction, with merging discs forming large
ellipticals (see e.g.\ \citealt{sch00} for a review).  Reconciliation
between two clearly distinct classes of galaxies, both in terms of
appearance and stellar population, with a transformation between the
two classes suggests that any such transformations are rapid, or
occurred at early times, in order to account for the relatively few
galaxies with intermediate properties, e.g., as in \
Figure~\ref{fig:pca_type}.

Numerical simulations show that the transformation from two disc
galaxies to a single merged remnant can occur in about 1.5 rotation
periods of the input discs, corresponding to $\sim$400Myr for galaxies
the size of the Milky Way (\citealt{sch00}), suggesting that
morphological transformations may indeed be fast enough to preserve a
bimodal population of galaxies.

\citet{bal04} investigate the transformation of galaxy colour as a
function of environment, finding similarly bimodal distributions.
Using \citet{bru03} synthetic stellar population models, they show
that if star-formation is ceased abruptly then galaxy colours
transform on the short time-scale of 750Myr.  If however,
star-formation decays with an exponential time-scale of 2Gyr, as
expected for a process such as strangulation, they show that colours
are transformed over a period of 4Gyr.

Note that, for massive cluster galaxies at least, morphological transformation
 must take place at a slower rate than transformation in
 star-formation rate, to account for the existence of a tight
 colour-magnitude relation for luminous early-type galaxies to high redshift
 (\citealt{sta98}; \citealt{hol04}), otherwise, at some redshift there
 would exist a population of bright blue elliptical galaxies which are not
 observed.  Note also that from our visual classifications we see very
 few such \emph{bright} blue ellipticals, but we do observe apparently quiescent
 spirals and \emph{faint} blue spheroids.  Thus transformation of the stellar populations in massive cluster galaxies must be
 occurring faster than $\sim 400$Myr if transformations are occurring
 at $z<1$.

Thus the bimodality of the galaxy populations could be due to rapid
transformation of both stellar-population and morphology, although
physically motivated mechanisms such as strangulation occur at slower
rates (\citealt{lar80}; \citealt{bal00}).

The bimodality could be produced through slower transformations, if
they occur mainly at high redshift.  There is indeed some evidence
that this is the case.  \citet{van99} finds a high rate of interaction
between galaxies in the $z=0.83$ cluster MS1054-0 3.  There is a close
link between mergers, ultra-luminous infrared galaxies and quasars
which suggests that the major epoch of merging occurred coincidentally
with the peak in quasar activity, around $z\approx2$
(\citealt{sch00}).  \citet{bun05} present evidence that morphological transformations of lower mass galaxies may take place at later cosmic epochs, which may explain why the majority of blue spheroids we see are low mass systems.

Note also that the major-merger rate, as measured in the MGC, is low
 (\citealt{dep05}), strongly suggesting that transformations as a
 result of mergers will not be very common today.  Also, \citet{lef00}
 see a strong increase in the fraction of major-mergers as a function
 of redshift in the Canada-France Redshift Survey.

These results suggest that the distinction between the two classes of
 galaxies is a result of the high redshift of the merging and
 transformation of galaxies.  We note however, that lack of merging at
 low redshifts is somewhat at odds with predictions of CDM, in which
 approximately 50\% of galaxies should have undergone a major merger
 since $z=1$ (\citealt{mur02}).

We postulate that a similar study of a high redshift sample of
galaxies, should produce different results and a blurring of the two
groups as the redshift of transformations and mergers is reached, if
transformations do indeed occur slowly at high redshift.

Finally we note that an analysis of the results of bulge-disc
decomposition and CAS (\citealt{con03}) analysis of the galaxies are
in progress and will be presented in future papers.

\section*{Acknowledgments}

The authors would like to thank Steve Lee for software help.  SCE
wishes to acknowledge PPARC support.

We thank Fionn Murtagh for making his hierarchical clustering software available.

The Millennium Galaxy Catalogue consists of imaging data from the
Isaac Newton Telescope and spectroscopic data from the Anglo
Australian Telescope, the ANU 2.3m, the ESO New Technology Telescope,
the Telescopio Nazionale Galileo, and the Gemini Telescope. The survey
has been supported through grants from the Particle Physics and
Astronomy Research Council (UK) and the Australian Research Council
(AUS). The data and data products are publicly available from
http://www.eso.org/$\sim$jliske/mgc/ or on request from J. Liske or
S.P. Driver.

Funding for the creation and distribution of the SDSS Archive has been
provided by the Alfred P. Sloan Foundation, the Participating
Institutions, the National Aeronautics and Space Administration, the
National Science Foundation, the U.S. Department of Energy, the
Japanese Monbukagakusho, and the Max Planck Society. The SDSS Web site
is http://www.sdss.org/.

The SDSS is managed by the Astrophysical Research Consortium (ARC) for
the Participating Institutions. The Participating Institutions are The
University of Chicago, Fermilab, the Institute for Advanced Study, the
Japan Participation Group, The Johns Hopkins University, the Korean
Scientist Group, Los Alamos National Laboratory, the
Max-Planck-Institute for Astronomy (MPIA), the Max-Planck-Institute
for Astrophysics (MPA), New Mexico State University, University of
Pittsburgh, University of Portsmouth, Princeton University, the United
States Naval Observatory, and the University of Washington.

\bibliographystyle{scemnras}
\bibliography{clusters}

\end{document}